%!TEX TS-program = pdftex

\def\V{{\cal V}}
\def\Rfi{{I\mskip-7mu R}}
\font\smallcaps=cmcsc10
\font\bfdue=cmbx12 scaled\magstep1
\magnification1200

{\voffset=6.8truecm
\hoffset=1.5truecm
\hsize=14truecm
\vsize=19truecm
\nopagenumbers
 
\centerline{{\bfdue  Complete Integrability}}

\medskip

\centerline{{\bfdue for Hamiltonian
                    Systems}}
\medskip
\centerline{{\bfdue with a Cone Potential}}
               
\bigskip                                                                    

\centerline{{\smallcaps Gianluca Gorni}}
\centerline{{\it Universit\`a di Udine}}
\centerline{{\it Dipartimento di Matematica 
             e Informatica}}
\centerline{{\it
             via Zanon 6,
             33100  Udine, Italy}}
\medskip
\centerline{{\smallcaps Gaetano Zampieri}}
\centerline{{\it Universit\`a di Padova}}
\centerline{{\it Dipartimento  di Matematica
        Pura e Applicata}}
\centerline{{\it
              via   Belzoni 7, 
              35131 Padova, Italy}}

\bigskip

\centerline{February 1989}

\bigskip
 
{\bf Abstract.}
It is known that, 
if a point in $\Rfi^n$ is driven
by a bounded below potential $\V$, 
whose gradient is always in a
closed convex cone which contains no lines, 
then the velocity has a finite
limit as time goes to $+\infty$.

The components of the asymptotic velocity, as
functions of the initial data, are trivially
constants of motion. We find sufficient conditions
for these functions  to be $C^k$ ($2\le k
\le+\infty$) first integrals, independent and
pairwise in involution.

In this way we construct a large class of
completely integrable systems.  We can deal
with very different asymptotic behaviours of the
potential and we have persistence of the
integrability under any small perturbation of the
potential in an arbitrary compact set.

\bigskip\vfil

\centerline{This research was supported by the
     {\it Ministero della Pubblica Istruzione}
            and by the C.N.R.}

\vskip1.6truecm

\eject
}

\font\bfuno=cmbx12
\font\bfdue=cmbx12 scaled\magstep1

\font\smallcaps=cmcsc10
\font\sf=cmss10
 
\def\Ham{{\cal H}} 
\def\V{{\cal V}}
\def\C{{\cal C}} 
\def\D{{\cal D}}
\def\F{{\cal F}}

\def\H{\,\hbox{\sf H}} 
\def\Dif{\,\hbox{\sf D}}
\def\Rfi{{I\mskip-7mu R}}

\def\sss{\scriptscriptstyle}

\pageno=3

\centerline{{\bfuno 1. 
                       Introduction}}
\bigskip

Given a smooth real function $(p,q)\mapsto H(p,q)$
defined in an open domain $\Omega$ of
$\Rfi^n\times\Rfi^n$ we can consider the 
associated  \it Hamiltonian system, \rm that is,
the autonomous system of ordinary differential
equations 
$$\dot q={\partial H\over\partial p}\,,\quad
  \dot p=-{\partial H\over\partial q}\,.
  \eqno(1.1)$$
The function $H$ is called a (time-independent)
Hamiltonian. We remind that the \it Poisson 
brackets \rm of two smooth real functions
$F,G\colon \Omega \to \Rfi$ are
$$\{F,G\}:=\sum_{i=1}^n\biggl(
  {\partial F\over\partial q_i}\, 
  {\partial G\over\partial p_i}\, -\, 
  {\partial F\over\partial p_i}\,
  {\partial G\over\partial q_i}\biggr)\,.
  \eqno(1.2)$$
An $F\in C^1(\Omega,\Rfi)$ 
is a constant of motion (or first integral)
for the system~(1.1) if and only if $\{F,H\}=0$.

Let us suppose that we find  $n$ functions
$F_1,\ldots,F_n\colon \Omega \to \Rfi$ of class 
$C^k$, $2\le k\le+\infty$, such that:

\medskip

\item{i)} $\{F_i,H\}=0$ for all $i$
   (i.e., the $F_i$ are first integrals of~(1.1));

\item{ii)} $\{F_i,F_j\}=0$ for all $i,j$
   (i.e., the $F_i$ are pairwise in
   involution);

\item{iii)} $\nabla F_1,\ldots,\nabla F_n$ are
   linearly independent in all of $\Omega$
   (the $F_i$ themselves are then said to be
   independent).

\medskip
\noindent
In this case a well known classical theorem says
that the system~(1.1) can be integrated by 
quadratures, in the usual sense of ordinary
differential equations (see~[2], Chapter~4,
Section~1.1).

If $H$ itself is one of the functions $F_i$ and
the solutions of the Hamiltonian systems
associated with each $F_i$ are all global (i.e.,
defined on $\Rfi$), then the system~(1.1) is
called  $C^k$-\it completely  integrable \rm
(see~[2], Chapter~4, Section~1.2). Analogously we
define  analytic integrability.

What is interesting about completely integrable
systems is that the  structure of the 
set of their solutions is very simple (see~[2],
Chapter~4, Section~1.2, Theorem~3). 
These properties, or rather the corresponding ones
in a more general setting, are the foundations of
a rich theory in the case when the level surfaces
of the vector function $(F_1,\ldots,F_n)$
are compact. This paper does not deal with this
last situation, but is concerned with proving the
complete integrability of some systems with
non-oscillatory behaviour, loosely related to
scattering problems.

\bigskip
\line{\hfil * \qquad * \qquad * \hfil }
\goodbreak
\bigskip
                                                 
Let $\V\colon\Rfi^n\to\Rfi$ be a smooth
function (called potential) and consider the
Hamiltonian  $\Ham(p,q):={1\over2}|p|^2+\V(q)$ 
with its associated system  
$$\dot q\, =\, p\,,\qquad\dot p\, 
  =\,-\nabla \V (q)\,. 
  \eqno(1.3)$$
Denote by 
$t\mapsto(p(t,\bar p,\bar q),q(t,\bar p,\bar q))$ 
the solution to~(1.3) with $(\bar p,\bar q)$ as 
initial data:
$$p(0,\bar p,\bar q)=\bar p\,,\quad
  q(0,\bar p,\bar q)=\bar q\,.$$

\goodbreak

Our starting point is the following assumption 
on the potential $\V$.

\bigskip

{\bf Hypothesis 1.1 } \sl $\V$ is a function in 
$ C^2 (\Rfi^n;\Rfi)$ such that:

\item{i)} $\V$ is bounded below;

\item{ii)}\sl there is a basis 
        $\{b_1,\ldots,b_n\}$ for $\Rfi^n$ 
        such that
        $ -\nabla \V (q)\cdot b_i\ge 0 $ 
        for all $ q\in \Rfi^n $ and all~$b_i$.
\rm  

\bigskip
   
Of course, the system admits the first integral of
energy 
$${1\over2}|p(t,\bar p,\bar q)|^2+
  \V(q(t,\bar p,\bar q))
  ={1\over 2}\,|\bar p|^2\,+\,\V(\bar q)\,.
  \eqno(1.4)$$
From i) we see that $|p(\cdot,\bar p,\bar q)|$ 
must be bounded for each solution, so that
by standard arguments in Ordinary 
Differential Equations 
we can prove that all solutions to~(1.3) are
defined for all times $t\in\Rfi$. 

On the other hand, property ii) implies that
$t\mapsto p(t,\bar p,\bar q)\cdot b_i$ is a
monotone function for all $(\bar p,\bar q)$ and
for all~$b_i$.

The whole of Hypothesis 1.1 thus ensures the
existence, along each solution, of the following
limit, the {\it asymptotic velocity:}
$$p_{\sss\infty}(\bar p,\bar q)
  :=\lim_{t\to +\infty} p(t,\bar p,\bar q)
    \in \Rfi^n\,.
    \eqno(1.5)$$
The limit as $ t\to -\infty $ exists as well.

These remarkably simple facts were pointed out by
Gutkin  in~[5]. He called the potentials $\V$
satisfying Hypo\-thesis~1.1~ii) {\it cone
potentials}. The reason for this name is as
follows. Let $\C$ be the convex cone in  $\Rfi^n$
spanned by the forces~$-\nabla\V$: $$ \C
:=\biggl\{ -\sum_{\alpha\in I}
  \lambda_{\alpha}\,\nabla \V (q_{\alpha})\;:
  \;\emptyset\ne I\hbox{ finite set, }
  \lambda_{\alpha}\ge0,\  q_{\alpha}\in \Rfi^n
  \;\forall\alpha\in I\biggr\}\, .
  \eqno(1.6)$$
and let $\D$ be the dual cone of $\C$, defined by
$$ \D:=\bigl\{w\in \Rfi^n\, :\,
   w\, \cdot \, v\, \ge\, 0\,
   \quad \forall
   v\in \C\bigr\}\,. \eqno(1.7)$$
Then Hypothesis~1.1 ii) means that $\D$ has
nonempty interior, or, equivalently, that the
closure of $\C$ contains no straight lines (such
cones $\C$ are called {\it proper}). We refer to
Section~2 for more details about cones.

\bigskip
\line{\hfil * \qquad * \qquad * \hfil }
\goodbreak
\bigskip

Let us survey
the content of the present paper. We are going to
provide only hints to our assumptions and results.
We will direct in each case to the precise
statements scattered through the following
Sections.

In Section~2 we give a few generalities about
cones in $\Rfi^n$ and prove a formula
(Pro\-position~2.4) that will be used extensively.

Section~3 presents three simple instances of cone
potentials for which the asymptotic velocity does
not depend continuously on the initial data. The
analysis of these counter\-examples leads in
Section~4 to write down our basic assumptions 
(the only {\it global} ones)% 
       {\parfillskip=0pt\par\goodbreak
        \parskip=0pt\noindent}%
on the potential $\V$, that, roughly speaking, 
amount to these:

\medskip
\nobreak

\item{1)} every level set of the potential $\V$ 
must be contained in a set of the form 
$q+\D$, so that the asymptotic velocity turn out
to belong to $\D$ (Hypo\-thesis~4.1);

\item{2)} the force $-\nabla\V$ must push
consistently toward the interior of~$\D\,$;
somewhat less roughly, the component of
$-\nabla\V(q)$ along any given direction of
$\bar\C$ shall be bounded below by a positive
constant, when $q$ varies on a (possibly
noncompact) set of a certain sort (Hypothesis~4.2).

\medskip

\noindent
Requirement 2) is actually the only severe
limitation for our approach. In particular, it
implies that $\C$ is contained in~$\D$, i.e., the
scalar product of any two vectors from~$\C$ is
nonnegative (i.e., $\C$ has width not larger than
$\pi/2$). Until Section~10 we will think of~$\V$
as being  defined on all of $\Rfi^n$, but
everything runs  just as well if $\V$ is defined
on a set of the form $q+\D^\circ$.

With the right hypotheses in hand, it becomes
easy to prove that the asymptotic velocity always
lies in the {\it interior} of the dual cone~$\D$
(Proposition~4.3), with certain \it locally
uniform \rm estimates on the trajectories
(Proposition~4.4). Such information is first used
in Section~5 to find general sufficient conditions
(Hypothesis~5.1) on the decay rate of $\V$ ``at
infinity'' (in the direction of the  cone~$\D$)
for $p_{\sss\infty}$ to be a continuous function
of the initial data. The tools are the fact that
$p_{\sss\infty}$~can be expressed as an integral: 
$$p_{\sss\infty}(\bar p,\bar q)=\bar p+
  \int_0^{+\infty}\mskip-14mu
  -\nabla\V(q(t,\bar p,\bar q))\,dt
  \eqno(1.8)$$
and the theorems on uniform integrability.

The first order differentiability of 
$p_{\sss\infty}$ is less immediate. We get it in
two different sets of assumptions. In Section~6 we
impose an {\it exponential} decay on the second
derivatives of~$\V$ (Hypothesis~6.1). This will
permit to exploit a simple Gronwall estimate on
the solutions of the  first variational equations
of our system,  and to use the theorems on
differentiation under the integral sign in~(1.8). 
In Section~7 we allow far more
general asymptotics for~$\V$, but we add the
side hypotheses of {\it convexity} on~$\V$ and a
kind of {\it monotonicity} in the Hessian matrix
(Hypothesis~7.1). A Liapunov function built on the
Hessian matrix of~$\V$ will give a sharp
control over the growth of the solutions of the 
first variational equation. As for the rest,
Sections~6 and~7 run very much parallel to each
other. Beside the mere regularity 
(Propositions~6.3 and~7.3), we also prove that
$p_{\sss\infty}$, as a function of the initial
data, is asymptotic, in the $C^1$ norm, to the
{\it projection}  $(p,q)\mapsto p$
(Propositions~5.3, 6.4 and~7.4). This will be
crucial in proving independence and involution in 
Section~9.

In Section~8 we show how to get higher order
differentiability of $p_{\sss\infty}$. This is not
difficult, since the bulk of the job has already
been done in Sections~6 and~7.

In Section~9 we reap the rewards of the regularity
theory to prove that the components of the
asymptotic velocity are first integrals,
independent and in involution, and to state the
{\it complete integrability} of  our systems
(Theorem~9.1). Furthermore, we show that the
potentials~$\V$ satisfying our sufficient
conditions for integrability can undergo arbitrary
(small enough) perturbations on any compact set of
$\Rfi^n$ without losing the property of yielding
completely integrable systems ({\it Persistence
Theorem~\rm9.2}). The fact that the integrability
is decided almost only on {\it asymptotic}
behaviour and survives generic modifications in a
bounded set  seems to be unusual in the theory of
integrable Hamiltonian systems.

\goodbreak

In Section~10 we give some examples. Namely, we 
provide manageable  conditions (Hypotheses~10.1)
on the functions $f_1,\ldots,f_N$ and on the
vectors $v_1,\ldots,v_N$ ($N\ge1$, no relation to
$n$) so that our theory applies to the
system with the potential~$\V$ given by 
$$\V(q):=\sum_{\alpha=1}^N
  f_\alpha(q\cdot v_\alpha)\,,\qquad
  q\in\Rfi^n
  \eqno(1.9)$$
(Proposition~10.5). A concrete instance is given
in Corollary~10.6: if $v_\alpha\cdot v_\beta\ge0$
for all $\alpha,\beta$ and if $r>0$, then the
Hamiltonian system with potential
$$\V(q):=\sum_{\alpha=1}^N{1\over(q\cdot
  v_\alpha)^r}\,,
  \qquad q\in\{\bar q\in\Rfi^n\,:\,\bar q\cdot
  v_\alpha>0
  \;\forall\alpha\}
  \eqno(1.10)$$
is $C^\infty$-completely integrable.

These cone potentials have
polyhedrical (that is, finitely generated) cone 
$\C$ of the forces (Lemma~10.2).  In a future
paper (in preparation) we will provide an
example where $\C$ is  {\it not polyhedrical}.
In~fact, the present approach  does not exploit
such additional  structures of~$\V$ as being
finite sum of one-dimensional functions. 

\medskip
\line{\hfil * \qquad * \qquad * \hfil }
\goodbreak
\medskip
 
An important analytically integrable system with
cone potential (and cone wider than $\pi/2$)
is the classical nonperiodic Toda Lattice system.
It describes the dynamics of $n$ particles on the
line  with coordinates $q_1,\ldots,q_n$
interacting through an exponential potential
$$ \V(q):=\sum_{i=1}^{n-1} e^{q_i-q_{i+1}}\,,
  \qquad q=(q_1,\ldots,q_n)\in\Rfi^n\,.
  \eqno(1.11)$$
This system was integrated by H\'enon and Flaschka
(with different methods, unrelated to cone
properties)---see the deep 
Lecture Notes~[11]. 

Subsequently, Gutkin in~[5] introduced cone
potentials, recognizing that a large subclass of
the  Toda-like potentials
$$\V(q):=\sum_{\alpha=1}^N c_\alpha 
  e^{q\cdot v_\alpha}\,,\qquad
  c_{\alpha}>0,\quad q\in \Rfi^n
  \eqno(1.12)$$
have the cone property (precisely,
if the convex cone $\bar\C$ spanned by the vectors
$v_1\ldots,v_N$ is proper). He noted that the
components of the asymptotic velocity, which
exists  for all systems with bounded below cone
potentials,  are likely candidates for being $n$
independent integrals in involution, as required
by the definition of integrability. However this
is not true  in general, since they may even be
discontinuous---see Section~3.

In the case when $\bar\C$ has amplitude not larger 
than $\pi/2$, the conjecture (namely,
$C^\infty$-integrability) was rigorously proved
by Oliva and Castilla in~[12] for the
potentials~(1.12) and also for~(1.9), but only
with functions $f_\alpha$ having exponential 
asymptotic behaviour at $+\infty$. Their method
uses the finite-sum form of the potential to
define a ``compactifying'' change of variables.
Then they prove and apply a Lemma in Dynamical
Systems (of independent interest too), concerning
the differentiability of a foliation of invariant
manifolds.%
       {\parfillskip=0pt\par\goodbreak
        \parskip=0pt\noindent}%
Oliva and Castilla drop the $\pi/2$
restriction for the following potentials: 
$$\eqalign{\V(q):=
  e^{q \cdot v_0}+e^{q \cdot v_{\theta}},\qquad
  &\theta \in \Rfi\,,\quad 
  q\in \Rfi^3,\quad v_{\sss0}=(1,-1,0),\cr 
  &v_{\theta}=\theta(0,1,-1)+
  (1-\theta)v_{\sss0} \cr} 
  \eqno(1.13)$$ 
with three  degrees of freedom, and 
$$\V(q):=e^{(-q_1-\alpha q_2)}+e^{(-q_1+\alpha
   q_2)},\qquad \alpha \ge 0\,, \quad
  q=(q_{\sss1},q_{\sss2})\in\Rfi^2
  \eqno(1.14)$$
with two freedoms,  and some generalizations
thereof,  all having either two or three freedoms.
We think that admitting wide cones needs such
strong restriction as low dimensions and/or
specialized  proofs exploiting the particular 
structure of a potential, and cannot be covered by
a  general theory of integrable systems with cone 
potential. The special role of the angle~$\pi/2$
is not so surprising if one thinks about the
behaviour of a billiard ball in a wedge. The
dynamics is fundamentally simpler if the wedge is 
wider than~$\pi/2$, and this corresponds to a
cone of the forces smaller than~$\pi/2$.

As for the potential~(1.14), Yoshida, Ramani, 
Grammaticos and Hietarinta~[13], using
Ziglin's methods~[14], proved that the associated
Hamiltonian system cannot be analytically
integrable if $\alpha^2\ne m(m-1)/2$ for $m$
integer.  Oliva and Castilla therefore remarked
that these systems are $C^\infty$ but not
analytically integrable.  For $0\le\alpha\le1$ 
the system defined by~(1.14) does fit into our
framework too (Corollary~10.7).  In particular,
\it we cannot  expect analytic integrability \rm
in the present approach either.
     
\medskip
\goodbreak

Another well-known cone potential (also with cone
wider than $\pi/2$) yielding an analytically
completely integrable system is 
$$\V(q):=\sum_{1\le i<j\le n}
  {1\over(q_i -q_j)^2}\,,\qquad q\in
  \{(q_1,\ldots,q_n)\in\Rfi^n\,:\,
  q_1<q_2<\ldots<q_n \}\,.
  \eqno(1.15)$$
It was introduced by Calogero and  Marchioro
(see~[3], [8], and~[4]) as the classical
counter\-part to a certain quantum mechanical
system. Marchioro proved (among many other
results) the integrability by explicit calculation
in the case  $n=3$. The integrability in the
general case was conjectured by Calogero and
proved by Moser~[10] through isospectral
deformations. 

Moauro, Negrini and Oliva [private communication] 
introduced the potentials~(1.10) with exponent
$r=2$ and put them into the framework of cone
potentials.  They proved the
$C^{\infty}$-integrability for $n=2$ and 3, even
when the cone $\bar\C$ has amplitude larger than
$\pi/2$. The compactification procedure, already
successful in~[12] for the exponential
case~(1.12), had to be supplemented with new ideas
to overcome the degeneracies arising in this
different situation. In particular, they use some
interesting techniques that had been developed
in~[9] in connection with a Liapunov Stability
problem.

\vfil

\noindent{\bf Acknowledgement.}  
We thank Piero Negrini for a critical discussion 
on the manuscript.

\vfil   \vfil\eject

%%%%%%%%%%%%%%%%%
\centerline{{\bfuno 2. Cones}}
\bigskip

{\bf Definition 2.1 } \sl A cone in $\Rfi^n$ is a
nonempty subset $C$ of $\Rfi^n$ such that
$v\in C$, $\lambda\ge0$
$\Rightarrow$ $\lambda v\in C$.
\rm

\bigskip

All the cones we will consider are convex, and
many of them closed, too. Note that the closure
of a (convex) cone is also a (convex) cone. 
We denote the closure of a subset $A$ of
$\Rfi^n$ either as $\bar A$ or as~cl$(A)$.

\bigskip

{\bf Definition 2.2 } \sl Given a convex cone
$C$ in $\Rfi^n$, the dual of $C$ is the set\rm
$$C^*:=\{w\in\Rfi^n\,:\, w\cdot v\ge0\;\;
  \forall v\in C\}\,.$$

\bigskip

It turns out that $C^*$ is a closed convex cone
and that $(C^*)^*=\bar C$.

\bigskip

{\bf Definition 2.3 } \sl A convex cone $C$ in
$\Rfi^n$ is called proper if its dual $C^*$ has
nonempty interior.\rm

\bigskip
\goodbreak

It is easy to see that a convex cone is proper
if and only if its closure contains no straight
lines. In fact, $\bar C$ contains a straight
line iff it contains both $v$ and $-v$ for some
$v\ne0$. And this is equivalent to $C^*\subset
\{w\,:\, w\cdot v=0\}$.

There is a neat way to express the distance
of a point of a convex cone from the boundary
in terms of the dual cone.
We will repeatedly use this formula.

\bigskip
\goodbreak

{\bf Proposition 2.4 } \sl Let $C$ be a 
convex cone in $\Rfi^n$, not reduced to $\{0\}$,
and let $D$ be its dual.
If $w\in D$ then \rm
$$\hbox{dist}(w,\partial D)=
  \min_{{v\in\bar C\atop|v|=1}}\; w\cdot v\,.$$

\bigskip
\goodbreak

{\bf Proof. } The distance of $w$ form
$\partial D$ is the same as the distance from 
$\hbox{cl}(\Rfi^n\backslash D)$, since $w\in D$.
On the other hand, $D$ is a closed convex
cone, so we have 
$$\hbox{cl}(\Rfi^n\backslash D)=
  \bigcup_{{v\in\bar C\atop|v|=1}}
    \{ u\in\Rfi^n\;:\; u\cdot v\le0\}\,.$$
In fact, from Hahn-Banach Theorem, for each
$u\in\hbox{cl}\,(\Rfi^n\backslash D)$, there is
an affine function $z\mapsto z\cdot v+a$, with
$|v|=1$, that is nonnegative on the convex $D$
and nonpositive on the half-line $\{\theta
u\,\colon\,\theta\ge0\}$. Since 0 belongs to both
sets, the constant $a$ is zero. Since $z\cdot
v\ge0$ for all $z\in D$, we have
$v\in D^*=(C^*)^*=\bar C$. This proves the
inclusion $\subset$. On the other hand, if
$u\in\Rfi^n$ and there exists $v\in\bar C$,
$|v|=1$, with $u\cdot v\le0$, then $u=\lim u_n$,
where $u_n\cdot v<0$. Hence 
$u_n\in\Rfi^n\backslash D$ and 
$u\in\hbox{cl}\,(\Rfi^n\backslash D)$.

\noindent We can write
$$\hbox{dist}(w,\partial D)=
  \inf_{{v\in\bar C\atop|v|=1}}\,
  \hbox{dist}\bigl( w\,,\,
         \{u\;\colon\;u\cdot v\le0\}\bigr)\,.$$
This last distance is the distance of $w$ from a
half-space. It thus coincides with $w\cdot v$.
The infimum is finally a minimum because the set
$\{v\in\bar C\,:\,|v|=1\}$ is compact and 
$v\mapsto w\cdot v$ is continuous.

\line{\hfill$\diamondsuit$}
\goodbreak
\bigskip

In the case of a polyhedral cone, the formula
becomes a minimum over a finite set. We will use
it in Section~10.

\bigskip

{\bf Proposition 2.5 } \sl Let $\bar C$ be the
closed cone generated by the vectors $v_1,\ldots,
v_N\in\Rfi^n\backslash\{0\}$. Then the dual cone
$D:=C^*$ is given by $D=\{w\,:\,
w\cdot v_\alpha\ge0\;\forall\alpha\}$ and for
all $w\in D$ we have\rm
$$\hbox{dist}(w,\partial D)=\min_\alpha
  w\cdot{v_\alpha\over|v_\alpha|}\,.$$

\bigskip
\goodbreak

{\bf Proof. } The formula for $D$ is easy.
Therefore $\Rfi^n\backslash D=
\bigcup_\alpha\{u\,:\,u\cdot v_\alpha<0\}$.
The union being finite, we can simply take
the closure and write 
$$\hbox{cl}(\Rfi^n\backslash D)=
  \bigcup_\alpha\{u\,:\,
  u\cdot v_\alpha\le0\}\,.$$
The rest of the proof is the same as for the
Proposition~2.4.

\line{\hfil$\diamondsuit$}
\goodbreak
\bigskip

{\bf Remark 2.6 } Let $D$ be a convex cone in
$\Rfi^n$,with nonempty interior and not
coincident with all of $\Rfi^n$. From Hahn-Banach
Theorem, we can separate the origin $0\in\Rfi^n$
from the interior of $D$, i.e., there exists
$v\in\Rfi^n\backslash\{0\}$ such that
$v\cdot w>0\quad\forall w\in D^\circ$.

  \vfil\eject

%%%%%%%%%%%%%%%%%

\centerline{{\bfuno 
             3. Counterexamples to Continuity}}
  
\bigskip

Consider the system~(1.3) for the 
following potentials.
   
\bigskip

{\bf Counterexample 3.1 } Let
$\V\colon\Rfi\to\Rfi\,,\quad q\mapsto e^{-q^3}$.

\medskip
   
Obviously, this is a  cone potential, that is, the
conditions of Hypothesis~1.1 hold. In fact 
$\V>0$ and $\bar\C=\Rfi_+\,$, where $\C$ 
is the cone generated by the forces---see (1.6). 
Moreover there is the equilibrium  $(p,q)=(0,0)$.

\noindent
For $\lambda > 0\,$, let us consider
$(p(\cdot,0,\lambda),q(\cdot,0,\lambda))$,
i.e., the solution of~(1.1) with $\bar p=0$ and
$\bar q=\lambda$ as initial conditions. Then 
$q(t,0,\lambda)\to +\infty$ as $t\to +\infty$ for
any $\lambda > 0\,$.

\noindent
The first integral of energy~(1.4) gives (for any
$t$)
$$|p(t,0,\lambda)|^2+\V\bigl(q(t,0,\lambda)\bigr)=
  e^{-\lambda^3}\,.$$
Therefore there is a discontinuity of
the asymptotic velocity $(\bar p,\bar q)\mapsto
p_{\sss\infty}(\bar p,\bar q)$  at
$(\bar p,\bar q)=(0,0)$ since
$|p_{\sss\infty}(0,\lambda)|^2\to 1$ as $\lambda
\to 0$ instead of $0=|p_{\sss\infty}(0,0)|^2$ as
for the equilibrium.

\line{\hfil$\diamondsuit$}
\goodbreak

\bigskip

In the previous counterexample  the lack of
continuity is related  with the presence of an
equilibrium. However, the absence of equilibria 
is not sufficient to guarantee  the continuity, 
as the following example shows.

\bigskip
\goodbreak

{\bf Counterexample 3.2 } Let 
$\V\colon\Rfi\to\Rfi\,,\quad 
 q\mapsto-\arctan q\,.$

\medskip
 
It is a cone potential because $\V>-1$ and 
$\bar\C=\Rfi_+\,$. Furthermore, there are no
equilibria. If the initial position is 
$\bar q=0\,$, and we conveniently choose
the initial velocity $\bar p<0$, then the
corresponding solution
$\bigl(p(t,\bar p,0),q(t,\bar p,0)\bigr)\to
(0,-\infty)$ as $t\to +\infty \,$. 
From the conservation of energy
$${1\over2}|p(t,\bar p,0)|^2-\arctan q(t,\bar p,0)
  ={1\over2}|\bar p|^2\,,$$
we see that the good choice for the initial
velocity is $ \bar p=-\sqrt\pi\,$.

\noindent
Now, let us reduce the initial speed, so that the
motion reverses its direction at a certain time. 
We easily see that the solution
$\bigl(p(\cdot,-\sqrt\pi+\lambda,0),
q(\cdot,-\sqrt\pi+\lambda,0)\bigr)$, 
for any $\lambda >0$, has the following  
asymptotic behaviour:
$$q(t,-\sqrt\pi+\lambda,0)\to+\infty\,,\qquad
  |p(t,-\sqrt\pi+\lambda,0)|^2\to 
  2\pi-2\lambda\sqrt\pi +\lambda^2
  \qquad
  \hbox{as } t\to+\infty\,.$$
The further limit as $\lambda\to0+$ proves the
discontinuity of  $(\bar p,\bar q)\mapsto
p_{\sss\infty}(\bar p,\bar q)$   at the point
$(\bar p,\bar q)=(-\sqrt\pi,0)$
(which are initial data of a solution with  
a vanishing asymptotic velocity, as we saw above).

\line{\hfill $\diamondsuit$}
\goodbreak
\bigskip

Of course the  preceding counterexample is 
possible because $\V$ does not go to $+\infty$ as 
$q\to-\infty$, i.e., because we do not have a \it
``barrier'' \rm in the direction opposite to the
force. 

\goodbreak

So far we have seen that we must avoid  the
equilibria and we need  a ``barrier'' in order 
that any motion eventually  ``points in the sense 
of the forces''. 
We shall give a precise formulation of  these
concepts in the next Section. Now let us give a
last example to show that the barrier is not yet 
sufficient. For this  we need two degrees of
freedom.

\bigskip

{\bf Counterexample 3.3 } 
$\V\colon \Rfi^2 \to \Rfi\,, \quad (q_1,q_2)
\mapsto  e^{-q_1^3}+e^{-q_2}\,.$

\medskip

This is a cone potential, 
$\{(v_{\sss1},v_{\sss2})\,:\,
 v_{\sss1}\ge0\,,\; v_{\sss2}>0\}\cup\{(0,0)\}$
being the cone $\C$ of the forces, and the dual
cone coinciding in this case too with $\bar\C$. 
We do not have equilibria.
The behaviour of the solutions can be easily  
investigated because the two degrees of 
freedom are separate. By Counterexample~3.1, we
see at once that a discontinuity in
$p_{\sss\infty}$ arises at the origin.

\line{\hfill$\diamondsuit$}
\bigskip

What seems to go wrong in the third counterexample
is that the force $-\nabla\V(q)$ does not drive
toward the {\it interior} of the dual cone for the
$q$ along the axis $q_{\sss1}=0$.

   \vfil\eject

%%%%%%%%%%%%%%%%%
\centerline{{\bfuno 4.
            Geometrical Bounds for the
            Asymptotic Velocity}}  
                                       
\bigskip
  
The basic ingredient of this work is the
potential function $\V$, about which we started
off with Hypo\-thesis~1.1. From $\V$ we
constructed the Hamiltonian function $\Ham$, the
Hamiltonian system~(1.3), its solutions 
$(p(t,\bar p,\bar q),q(t,\bar p,\bar q))$, the
asymptotic velocity 
$p_{\sss\infty}(\bar p,\bar q)$, the convex
cone $\C$ spanned by the forces~(1.6) and its 
dual $\D:=\C^*$~(1.7).

The next assumptions on $\V$ are the first steps
toward integrability.

\bigskip
 
{\bf Hypothesis 4.1\ }\ {\sl For each $M>0$ 
there exists a $q_{\sss M}\in\Rfi^n$ such that}
$$q\in \Rfi^n\backslash
  (q_{\scriptscriptstyle M}+\D)
  \quad\Rightarrow\quad 
  \V(q)\ge M .$$

\bigskip
\goodbreak

{\bf Hypothesis 4.2 } \sl For each 
$q^\prime\,,\,q^{\prime\prime}\in\Rfi^n$ 
such that 
$q^{\prime\prime}\in  q^\prime+\D$,
and for each
$v\in\bar\C\backslash\{0\}$ 
there exists 
$\varepsilon>0$ such that
$$\Bigl(\; q\in q^\prime+\D
  \quad\hbox{and}\quad 
  q\cdot v\le
  q^{\prime\prime}\cdot v \;\Bigr)
  \quad\Rightarrow\quad
  -\nabla\V(q)\cdot v\ge\varepsilon\,.$$
\rm

\bigskip
\goodbreak

Hypothesis 4.2 implies in particular that 
$-\nabla\V(q)\cdot v>0$ for all $q\in\Rfi^n$,
and all $v\in\bar\C\backslash\{0\}$. Therefore
$\C\subset\D$ and $\C$ has amplitude not
larger than $\pi/2$.

\bigskip
\goodbreak

{\bf Proposition 4.3 } \sl If
Hypo\-theses~1.1, 4.1 and~4.2 hold, then 
$$p_{\sss\infty}(\bar p,\bar q)\cdot v>0  
  \eqno(4.1)$$ 
for all $v\in\bar\C\backslash\{0\}$  and all
initial data $(\bar p,\bar q)$. This is the same
as saying that $p_{\sss\infty}(\bar p,\bar q)$
lies in the interior of the dual cone $\D$.\rm

\bigskip
\goodbreak

{\bf Proof. } Let us fix the initial data 
$(\bar p,\bar q)$.
The potential is
bounded above along the trajectory:
$$\V(q(t,\bar p,\bar q))\le{1\over2}|\bar p|^2+
  \V(\bar q)\,:=M \qquad \forall t\in\Rfi
  \eqno(4.2)$$
by the conservation of energy. Hypo\-thesis~4.1
alone guarantees  that $q(t,\bar p,\bar q)$
remains in $q_{\sss M}+\D$ for all times.
This already implies that
$p_{\sss\infty}(\bar p,\bar q)$ belongs  to the
closed set $\D$.

\noindent Let $v\in\bar\C\backslash\{0\}$. 
As we remarked immediately after Hypothesis 4.2,
$$\dot p(t,\bar p,\bar q)\cdot v 
  =   -\nabla\V(q(t,\bar p,\bar q))\cdot v 
  > 0 $$
and so $t\mapsto p(t,\bar p,\bar q)\cdot v$ is an
increasing funtion. 

\noindent We argue by contradiction.
If $p_{\sss\infty}(\bar p,\bar q)\cdot v$ 
happened to be nonpositive, then 
$$\dot q(t,\bar p,\bar q)\cdot 
  v=p(t,\bar p,\bar q)\cdot v
  < 0 
  \qquad \forall t\in\Rfi$$
and $t\mapsto q(t,\bar p,\bar q)\cdot v$ 
would be decreasing. Hence
$$t\ge0 \quad\Rightarrow\quad
  q(t,\bar p,\bar q)\cdot v \le
  \bar q\cdot v \,.$$ 
Hypothesis 4.2 now yields that for $t\ge0$ the
scalar product  $-\nabla\V(q(t,\bar p,\bar q))
\cdot v$ is not less than some $\varepsilon>0$.
Thus we can write: $$\eqalign{
     p(t,\bar p,\bar q)\cdot v
         &=  \bar p\cdot v +
            \int_0^t-\nabla\V(q(s,\bar p,\bar q))
                     \cdot v\,ds\ge       \cr
         &\ge \bar p\cdot v +
            \varepsilon t \quad
                 \to +\infty \quad 
                 {\rm as}\  t\to+\infty\,, \cr}$$
and this contradicts the assumption 
$p_{\sss\infty}(\bar p,\bar q)\cdot v\le0$.
Finally, formula~(4.1) is equivalent to
$p_{\sss\infty}(\bar p,\bar q)\in\D^\circ$
because of Proposition~2.4.

\line{\hfill $\diamondsuit$}
\goodbreak
\bigskip

In the sequel we will need the following
information about the trajectories: locally
uniformly in the initial data, 

\medskip

\item{1)} the velocity $p(t,\bar p,\bar q)$
is eventually in the interior of the dual cone,
its distance from the boundary remains larger
than a positive number $\gamma$, and 
\item{2)} the position $q(t,\bar p,\bar q)$
enters and no longer quits
any set of the form $q_{\sss0}+\D$.

\goodbreak
\bigskip

{\bf Proposition 4.4 } \sl In the hypotheses of
Proposition 4.3, for each $(\bar p_{\sss0},
\bar q_{\sss0})\in\Rfi^n\times\Rfi^n$ and each
$q_{\sss0}\in\Rfi^n$ there esist $\gamma>0$,
$t_{\sss0}\in\Rfi$ and a bounded neighbourhood $U$
of  $(\bar p_{\sss0},\bar q_{\sss0})$ in 
$\Rfi^n\times\Rfi^n$ such that, for all 
$t\ge t_{\sss0}$ and $(\bar p,\bar q)\in U$ we
have \rm
$$\matrix{p(t,\bar p,\bar q)\in\D^\circ,\hfill&
  \hbox{dist}\bigl(p(t,\bar p,\bar q),\partial\D
  \bigr)\ge\gamma\,,\hfill\cr
  q(t,\bar p,\bar q)\in q_{\sss0}+\D\,,\hfill
  &\hbox{dist}\bigl(q(t,\bar p,\bar q),q_{\sss0}
  +\partial\D\bigr)\ge
  \gamma(t-t_{\sss0})\,.\hfill
  \cr}$$

\bigskip
\goodbreak

{\bf Proof. } Since $p_{\sss\infty}(\bar 
p_{\sss0},\bar q_{\sss0})\in\D^\circ$, let
$\gamma:=(1/2)\hbox{dist}\bigl(p_{\sss\infty}
(\bar p_{\sss0},\bar q_{\sss0}),\partial\D
\bigr)>0$. Because of Proposition~2.4 and the
continuity of the distance function, there exist 
$t_{\sss1}\in\Rfi$ and a bounded neighbourhood 
$U$ of $(\bar p_{\sss0},\bar q_{\sss0})$ such
that 
$$p(t_{\sss1},\bar p,\bar q)
  \in\D^\circ,\qquad
  \hbox{dist}\bigl(p(t_{\sss1},\bar p,\bar q),
  \partial\D\bigr)\ge\gamma$$
for all $(\bar p,\bar q)\in U$. But for all 
$(\bar p,\bar q)$ and all $v\in\bar\C\backslash
\{0\}$ the function $t\mapsto p(t,\bar p,\bar q)
\cdot v$ is increasing, so that the velocity
$p(t,\bar p,\bar q)$ lies in $\D^\circ$ for all
$t\ge t_{\sss1}$ and all $(\bar p,\bar q)\in U$,
and its distance from the boundary is not less
than $\gamma$.
For all $(\bar p,\bar q)\in U$, $t\ge
t_{\sss1}$, $v\in\bar\C$, $|v|=1$,
we have
$$\eqalign{
  \bigl(q(t,\bar p,\bar q)-q_{\sss0}\bigr)
  \cdot v&=
  \bigl(q(t_{\sss0},\bar p,\bar q)-q_{\sss0}
  \bigr)\cdot v+\int_{t_1}^t p(s,\bar p,\bar q)
  \cdot v\,ds\ge\cr
  &{}\ge
  \inf_{(\bar p,\bar q)\in U}
  \inf_{{w\in\bar\C\atop|w|=1}}
  \bigl(q(t_{\sss1},\bar p,\bar q)-q_{\sss0}
  \bigr)\cdot w +\gamma(t-t_{\sss1})
  :=a+\gamma (t-t_{\sss1})\,,\cr}$$
and finally, for 
$t\ge t_{\sss0}:=\max\{t_{\sss1}, t_{\sss1}-(a/
\gamma )\}$ the point $q(t,\bar p,\bar q)$
belongs to $q_{\sss0}+\D$ and
$$\hbox{dist}\bigl(q(t,\bar p,\bar q),
  q_{\sss0}+\partial\D\bigr)=
  \min_{{v\in\bar\C\atop|v|=1}}
  \bigl(q(t,\bar p,\bar q)-q_{\sss0}\bigr)
  \cdot v\ge\gamma (t-t_{\sss0})\,.$$

\line{\hfill$\diamondsuit$}
\goodbreak
\bigskip

We may ask what happens of $\V$ and $\nabla\V$
along the trajectories.

\bigskip

{\bf Proposition 4.5 } \sl In the hypotheses of
Proposition~4.3, for all initial data 
$(\bar p,\bar q)\in\Rfi^n\times\Rfi^n$ we have\rm
$$\lim_{t\to+\infty}\V(q(t,\bar p,\bar q))=
  \inf\V\,.$$

\bigskip
\goodbreak

{\bf Proof. } Fix $\varepsilon>0$ and pick
$q_\varepsilon\in\Rfi^n$ such that 
$\V(q_\varepsilon)\le\inf\V+\varepsilon$.
Let $q\in q_\varepsilon+\D$. Then
$$\eqalign{
  \V(q)-\V(q_\varepsilon)&=
  \int_0^1{d\over d\theta}\V(q_\varepsilon+
  \theta(q-q_\varepsilon))\,d\theta=\cr
  &{}=\int_0^1\nabla\V(q_\varepsilon+
  \theta(q-q_\varepsilon))\cdot
  (q-q_\varepsilon)\,d\theta\le0\,,\cr}$$
because $-\nabla\V(q_\varepsilon+
\theta(q-q_\varepsilon))\in\C$ and
$q-q_\varepsilon\in\D=\C^*$.
Hence we can write
$$q\in q_\varepsilon+\D\quad\Rightarrow\quad
  \V(q)\le\inf\V+\varepsilon\,.$$
On the other hand, Proposition~4.4 guarantees,
in particular, that for all $(\bar p,\bar q)$
there exists $t_\varepsilon\in\Rfi$ such that
$$t\ge t_\varepsilon\quad\Rightarrow\quad
  q(t,\bar p,\bar q)\in q_\varepsilon+\D\,.$$
This concludes the proof.

\line{\hfill$\diamondsuit$}
\goodbreak
\bigskip

{\bf Corollary 4.6 }\sl In the hypotheses of
Proposition~4.3, the following identity holds:
\rm
$$\Ham(\bar p,\bar q)={1\over2}|\bar p|^2+
  \V(\bar q)=
  {1\over2}|p_{\sss\infty}(\bar p,\bar q)|^2+
  \inf\V\,.$$

\bigskip

Within the assumptions of this Section, the
gradient $\nabla\V$ does not need to be
infinitesimal along the trajectories. Already in
one dimension, it is not difficult to figure out
a $\V\in C^2(\Rfi,\Rfi)$ that decreases from
$+\infty$ to~0, and whose graph has infinitely
many smooth, but steep steps (whose height will
obviously tend to zero):
$$\inf\V=0\,,\quad \sup\V=+\infty\,,\quad
  \V^\prime<0\,,\quad
  \liminf_{q\to+\infty}\V^\prime(q)<0\,.$$
All the $q(t)$ will go to $+\infty$ as
$t\to+\infty$ because our hypotheses are
verified, so that they will never stop undergoing
jerks ($\nabla\V$ does not converge). 
The assumptions of the next Section will rule out
this possibility.

   \vfil\eject

%%%%%%%%%%%%%%%%%

\centerline{{\bfuno 5. Continuity}}
\bigskip

This Section deals with the continuity of the
asymptotic velocity with respect to the initial
data. Gutkin in~[7] already studied the
problem, but in his setting he had no guarantee
that $p_{\sss\infty}$ belonged to the interior of
the dual cone for all the trajectories. Much less
did he obtain such crucial estimates as the
ones in Proposition~4.4. So he obtained
the continuity in a nonempty, open subset of the
space of the initial data, defined in terms of
$p_{\sss\infty}$ itself. 

In our assumptions, we get {\it global}
continuity. We will also prove an asymptotic
property of $p_{\sss\infty}\,$, that will enable
us later to determine the exact range of the
mapping~$p_{\sss\infty}$ (Proposition~7.5).

The asymptotic velocity 
$p_{\sss\infty}(\bar p,\bar q)=\lim_{t\to+\infty}
 p(t,\bar p,\bar q)$
can be expressed in terms of an integral:
$$p_{\sss\infty}(\bar p,\bar q)=\bar p +
  \int_0^{+\infty}\mskip-14mu
  -\nabla\V(q(s,\bar p,\bar q))
  \,ds\,.\eqno(5.1)$$
In the hypotheses of the last section, we know
that $q(\cdot,\bar p,\bar q)$ is always
contained in $q_{\sss M}+\D$ (see formula~(4.2)).
Moreover,  
$p_{\sss\infty}(\bar p,\bar q)$ is in the
{\it interior} of the dual cone $\D$, so that
the distance of $q(t,\bar p,\bar q)$ from the
boundary of $q_{\sss M}+\D$ grows
linearly as $t\to+\infty$, as we saw in
Proposition~4.4.
We may expect $p_{\sss\infty}$
to be a continuous function of $(\bar p,\bar q)$
if the norm $|\nabla\V(q)|$ is dominated
by an integrable function of the distance 
between $q$ and $q_{\sss M}+
\partial\D$.
We may thus use a uniform
integrability theorem on the integral~(5.1).

\bigskip
\goodbreak

{\bf Hypothesis 5.1 } \sl There exist
$q_{\sss0}\in\Rfi^n$ and an
weakly decreasing, integrable function
$h_{\sss0}\colon\Rfi_+\to\Rfi$  such that \rm
$$q\in q_{\sss0}+\D
  \quad\Rightarrow\quad
  |\nabla\V(q)|\le
  h_{\sss0}\Bigl(\,
    \hbox{dist}\bigl(q,q_{\sss0}
                    +\partial\D\bigr)\,
  \Bigr)\,.$$

\bigskip
\goodbreak

{\bf Proposition 5.2 } \sl If Hypotheses~1.1,
4.1, 4.2 and~5.1 hold, then $p_{\sss\infty}$ is a
continuous function of the initial data. \rm

\bigskip
\goodbreak

{\bf Proof. } Let $(\bar p_{\sss0},\bar
q_{\sss0})$ be fixed and pick $\gamma >0$,
$t_{\sss0}\in\Rfi$ and $U$ from Proposition~4.4:
$$q(t,\bar p,\bar q)\in q_{\sss0}+\D\,,
  \qquad\hbox{dist}\bigl(
  q(t,\bar p,\bar q),q_{\sss0}+\partial\D
  \bigr)\ge\gamma (t-t_{\sss0})$$
for all $t\ge t_{\sss0}$, $(\bar p,\bar q)\in U$.
Using Hypo\-thesis~5.1,
$$|\nabla\V(q(t,\bar p,\bar q))|\le
  h_{\sss0}\Bigl( \hbox{dist}\bigl
  (q(t,\bar p,\bar q),
  q_{\sss0}+\partial\D\bigr)\Bigr)
  \le
  h_{\sss0}\bigl(\gamma (t-t_{\sss0})\bigr)\,,$$ 
so that we can apply the theorems on uniform 
integrability to the formula 
$$p_{\sss\infty}(\bar p,\bar q)=
  p(t_{\sss0},\bar p,\bar q)+
  \int_{t_{\sss0}}^{+\infty}\mskip-14mu
  -\nabla\V(q(s,\bar p,\bar q))\, ds$$
and obtain our continuity result.

\line{\hfill $\diamondsuit$}
\goodbreak
\bigskip

We are now provided with $n$ continuous integrals
of motion: the components of the asymptotic
velocity $p_{\sss\infty}(\bar p,\bar q)$.

Roughly speaking, if we find a region $q+\D$
where the driving force $-\nabla\V$ is utterly
negligible, we may expect that, if we start the
motion there, with a velocity $\bar p$ in the
interior of $\D$, then the motion has
approximately constant speed: $$p(t,\bar p,\bar
q)\approx \bar p$$ so that, for those initial
data  $(\bar p,\bar q)$ we have
$$p_{\sss\infty}(\bar p,\bar q)
  \approx\bar p\,.\eqno(5.3)$$

\bigskip
\goodbreak

{\bf Proposition 5.3 } \sl In the hypotheses of
Proposition~5.2, for each $\mu>0$ and each
$\gamma >0$, there exists
$q_{\sss0}^\prime\in\Rfi^n$ such that \rm
$$\Bigl(\; \bar p\in\D^\circ\,,\quad
  \hbox{dist}(\bar p,\partial\D)\ge\gamma \,,
  \quad\bar q\in q_{\sss0}^\prime
  +\D \;
  \Bigr)\quad\Rightarrow\quad
  \bigl| p_{\sss\infty}(\bar p,\bar q)-\bar p
  \bigr|\le\mu\,.$$

\bigskip
\goodbreak

{\bf Proof. } Let $\mu>0$, $\gamma >0$ be fixed
and pick $q_{\sss0}$ from Hypo\-thesis~5.1. Let
$d_{\sss0}\ge0$ be such that
$$\int_{d_{\sss0}}^{+\infty}\mskip-14mu
  h_{\sss0}(\gamma  t)\,dt\le\mu\,.$$
Choose $q_{\sss0}^\prime\in q_{\sss0}+\D$ 
such that  $\hbox{dist}(q_{\sss0}^\prime,
q_{\sss0}+\partial\D)\ge d_{\sss0}$. 
Then, for all $\bar p\in\D^\circ$ such
that $\hbox{dist}(\bar p,\partial\D)\ge\gamma $
and $\bar q\in q_{\sss0}^\prime+\D$ we have
$$\hbox{dist}\bigl(
  q(t,\bar p,\bar q),q_{\sss0}+\partial\D\bigr)
  \ge\gamma  t+d_{\sss0}$$
and so
$$\eqalign{\bigl| p_{\sss\infty}(\bar p,\bar q)
  -\bar p\bigr|&{}=
  \biggl|\int_0^{+\infty}\mskip-14mu
  -\nabla\V(q(t,\bar p,\bar q))\,dt\biggr|\le\cr
  &{}\le\int_0^{+\infty}\mskip-14mu
  h_{\sss0}(\gamma  t+d_{\sss0})\,dt=
  \int_{d_{\sss0}}^{+\infty}\mskip-14mu
  h_{\sss0}(\gamma  t)\,dt\le\mu\,.\cr}$$

\line{\hfil$\diamondsuit$}
\goodbreak
 
  \vfil\eject

%%%%%%%%%%%%%%%%%
\centerline{{\bfuno 6. 
             First Order Differentiability}}
\smallskip
\centerline{{\bfuno without Convexity on the
             Potential}}
\bigskip

In the Hypotheses 1.1 we know that the velocity
$p(t,\bar p,\bar q)$ is a differentiable
function of the initial data at all finite times
$t$. If we differentiate the equation
$$p(t,\bar p,\bar q)=p(t_{\sss0},\bar p,\bar q)+
  \int_{t_{\sss0}}^t -\nabla\V(q(s,\bar p,
  \bar q))\,ds
  \eqno(6.1)$$
with respect to an arbitrary component of $\bar p$
or $\bar q$, we obtain
$$Dp(t,\bar p,\bar q)=Dp
  (t_{\sss0},\bar p,\bar q)+
  \int_{t_{\sss0}}^t -\H\V(q(s,\bar p,\bar q))
  Dq
  (s,\bar p,\bar q)\,ds\,,
  \eqno(6.2)$$
where we denote by $\H\V$ the Hessian matrix of
$\V$ and by $D$ the partial derivative (we
reserve the character $\Dif$ for the Jacobian
matrix). 

In the hypotheses of Section~4, we know that
formula~(6.1) holds with $t=+\infty$ and the
integrability is uniform. Can we expect the same
for~(6.2)? What we seem to need is:
\medskip
\item{1)} an a priori bound on the growth of 
  $Dq(t,\bar p,\bar q)$, locally
  uniform on $(\bar p,\bar
  q)$;
\item{2)} a rapid enough decrease of the norm of
  the Hessian $\H\V$  along the trajectories 
  $t\mapsto q(t,\bar p,\bar q)$.
\medskip

\noindent If the two estimates match
appropriately, we can use the theorems on the
differentiation under the integral sign.

\goodbreak

The function $z(t)=Dq(t,\bar p,\bar q)$ satisfies
the linear differential equation
$$\ddot z(t)=-\H\V(q(t,\bar p,\bar q))\,z(t)\,,
  \eqno(6.3)$$
that can also be rewritten as a first-order
system:
$${d\over dt}{z(t)\choose \dot z(t)}
  =\left(\matrix{0& I_n\cr
  -\H\V(q(t,\bar p,\bar q))&0\cr}\right)
  {z(t)\choose \dot z(t)}\,.$$
Let us denote by $R(t,s,\bar p,\bar q)$ the
evolution operator of the system, i.e., the $2n
\times 2n$ matrix solution of
$${\partial\over\partial t}R(t,s,\bar p,\bar q)=
  \left(\matrix{0& I_n\cr
  -\H\V(q(t,\bar p,\bar q))&0\cr}\right)
  R(t,s,\bar p,\bar q)\,,\qquad
  R(s,s,\bar p,\bar q)=I_{2n}\,.$$
Let $\Pi$ and $\Pi^\prime$ be the two projections
$\Rfi^n\times\Rfi^n\to\Rfi^n$ defined as
$\Pi(x,y)=x$, $\Pi^\prime(x,y)=y$. Since 
$$Dq(t,\bar p,\bar q)=\Pi R(t,0,\bar p,\bar q)
  {Dq(0,\bar p,\bar q)\choose
  Dp(0,\bar p,\bar q)}
  =\Pi R(t,0,\bar p,\bar q){D\Pi^\prime
  \choose D\Pi},$$
what we are interested in is the behaviour of 
$\|\Pi R(t,s,\bar p,\bar q)\|$ as $t\to+\infty$.

\goodbreak

We will carry out this program in two sets of
hypotheses. In the rest of this Section our
assumptions will be as follows.

\bigskip

{\bf Hypothesis 6.1 } \sl There exist
$q_{\sss1}\in\Rfi^n$, $A_{\sss1}\ge0$,
$\lambda_{\sss1}>0$ such that \rm
$$q\in q_{\sss1}+\D\quad\Rightarrow\quad
  \|\H\V(q)\;\|\le
  A_{\sss1}\exp\Bigl(-\lambda_{\sss1}\hbox{
  dist}\bigl(
  q,q_{\sss1}+\partial\D\bigr)\,\Bigr)\,.$$

\bigskip
\goodbreak

On one hand, the mere fact that the Hessian is
infinitesimal along the trajectories $t\mapsto
q(t,\bar p,\bar q)$ ensures, via a Gronwall
lemma, that  $Dq(t,\bar p,\bar q)$ must grow 
{\it less than exponentially} as  $t\to+\infty$
(i.e., it is $o(e^{\varepsilon t})$ for all
$\varepsilon>0$).  On the other hand the actual
{\it exponential} decrease of the Hessian
compensates for the other growth and yields the
uniform integrability of~(6.2).
 
\bigskip
\goodbreak

{\bf Lemma 6.2 } \sl 
Suppose that Hypo\-theses~1.1, 4.1, 4.2 and~6.1
hold. Then, for all $\varepsilon>0$, and for all
$t\ge0$, $x$, $y\in\Rfi^n$ we have \rm
$$\eqalign{
  \biggl(
  \bar p\in\D\,,&\quad \bar q\in q_{\sss1}+\D\,,
  \quad \hbox{dist}(\bar q,q_{\sss1}+\partial
  \D)\ge-{1\over\lambda_{\sss1}}
  \ln{\varepsilon^2\over 
  A_{\sss1}}\biggr)\quad\Rightarrow\cr
  &\Rightarrow\quad
  \Bigl| \Pi R(t,0,\bar p,\bar q)
  {x\choose y}\Bigr|\le
  {1\over2\varepsilon}\Bigl(
  (\varepsilon|x|+|y|)e^{\varepsilon t}+
  (\varepsilon|x|-|y|)e^{-\varepsilon t}
  \Bigr)\,.\cr}$$

\bigskip
\goodbreak

{\bf Proof. } Choose $\varepsilon>0$, $\bar q\in
q_{\sss1}+\D$ such that $\hbox{dist}(\bar
q,q_{\sss1}+\partial\D)\ge
(-1/\lambda_{\sss1})\ln(\varepsilon^2/A_{\sss1})$.
For any $\bar p\in\D$, we have
$$\hbox{dist}\bigl(q(t,\bar p,\bar q),
  q_{\sss1}+\partial\D\bigr)\ge
  \hbox{dist}(\bar q,q_{\sss1}+\partial\D)$$
and so
$$\eqalign{
  \bigl\|\H\V(q(t,\bar p,\bar q))\bigr\|&\le
  A_{\sss1}\exp\Bigl(-\lambda_{\sss1}\,
  \hbox{dist}\bigl(
  q(t,\bar p,\bar q),q_{\sss1}+\partial\D
  \bigr)\Bigr)\le\cr
  &{}\le A_{\sss1}\exp\Bigl(
  -\lambda_{\sss1}\,\hbox{dist}
  \bigl(\bar q,q_{\sss1}+\partial\D\bigr)\Bigr)
  \le\varepsilon^2\,.\cr}$$
For any $x$, $y\in\Rfi^n$, the function
$z(t)=\Pi R(t,0,\bar p,\bar q){x\choose y}$ is a
solution of~(6.3), that can be rewritten in
integral form:
$$z(t)=x+ty+\int_0^t-(t-s)
  \H\V(q(s,\bar p,\bar q))\,z(s)\,ds\,.$$
Taking the norms,
$$\eqalign{
  |z(t)|&\le|x|+t|y|+\int_0^t(t-s)
  \bigl\|\H\V(q(s,\bar p,\bar q))\bigr\|\,
  |z(s)|\,ds\le\cr
  &{}\le|x|+t|y|+
  \varepsilon\int_0^t(t-s)|z(s)|\,ds\,.\cr}$$
A standard Gronwall argument yields that 
$|z(t)|\le\varphi(t)$ for $t\ge0$, where 
$\varphi$ is the solution of
$$\varphi(t)=|x|+t|y|+\varepsilon\int_0^t
  (t-s)\varphi(s)\,ds\,,$$
and is precisely the expression appearing in
the statement of the Lemma.

\line{\hfil$\diamondsuit$}
\goodbreak
\bigskip

{\bf Proposition 6.3 } \sl Suppose that
Hypo\-theses~1.1, 4.1, 4.2, and~6.1 hold. Then
the asymptotic velocity  $p_{\sss\infty}$ is a
$C^1$ function of the initial data.\rm

\bigskip
\goodbreak

{\bf Proof. } What we need is local uniform
integrability of the integral in~(6.2) for some
$t_{\sss0}\in\Rfi$. Choose an initial condition
$(\bar p_{\sss0},\bar q_{\sss0})$. From
Proposition~4.4, there exist $\gamma >0$,
$t_{\sss1}\in\Rfi$ and a bounded neighbourhood $U$
of $(\bar p_{\sss0},\bar q_{\sss0})$ such that
$$\matrix{
  p(t,\bar p,\bar q)\in\D^\circ,\hfill&
  \hbox{dist}\bigl(p(t,\bar p,\bar q),
  \partial\D\bigr)\ge\gamma \,,\hfill\cr
  q(t,\bar p,\bar q)\in q_{\sss1}+\D\,,\hfill&
  \hbox{dist}\bigl(q(t,\bar p,\bar q),
  q_{\sss1}+\partial\D\bigr)\ge
  \gamma (t-t_{\sss1})\hfill\cr}$$
for all $t\ge t_{\sss1}$, $(\bar p,\bar q)\in U$.
So we have
$$\bigl\|\H\V(q(t,\bar p,\bar q))\bigr\|\le 
  A_{\sss1}e^{-\lambda_{\sss1}\gamma(t-t_1)}$$
for all $t\ge t_{\sss1}$, $(\bar p,\bar q)\in U$.
Now choose $\varepsilon>0$ and $t_{\sss0}\ge
t_{\sss1}$ such that
$$0<\varepsilon<\lambda_{\sss1}\gamma 
  \quad\hbox{and}\quad
  A_{\sss1} e^{-\lambda_{\sss1}\gamma (t_0-t_1)}
  \le\varepsilon^2\,.$$
Since
$$Dq(t,\bar p,\bar q)=\Pi R(s,0,\bar p,\bar q)
  {D\Pi^\prime\choose D\Pi}=
  \Pi R\bigl(s-t_{\sss0},0,
  p(t_{\sss0},\bar p,\bar q),
  q(t_{\sss0},\bar p,\bar q)\bigr)
  {Dq(t_{\sss0},\bar p,\bar q)\choose
  Dp(t_{\sss0},\bar p,\bar q)},
  \eqno(6.4)$$
from Lemma~6.2 we get that
$$\bigl|Dq(t,\bar p,\bar q)\bigr|\le
  a_{\sss1}e^{\varepsilon(t-t_0)}+a_{\sss2}
  \eqno(6.5)$$
for all $t\ge t_{\sss0}$ and all $(\bar p,\bar q)
\in U$, where
$$a_{\sss1}\,:=\sup_{(\bar p,\bar q)\in U}
  {1\over2\varepsilon}\Bigl(
  \varepsilon|Dq(t_{\sss0},\bar p,\bar q)|
  +|Dp(t_{\sss0},\bar p,\bar q)|
  \Bigr),\qquad
  a_{\sss2}\,:=\sup_{(\bar p,\bar q)\in U}
  {1\over2}|Dq(t_{\sss0},\bar p,\bar q)|\,.$$
We can finally write, for all $t\ge t_{\sss0}$,
$(\bar p,\bar q)\in U$:
$$\bigl|-\H\V(q(t,\bar p,\bar q))
  Dq(t,\bar p,\bar q)\bigr|\le
  A_{\sss1}e^{-\lambda_{\sss1}\gamma (t-t_1)}
  (a_{\sss1}e^{\varepsilon(t-t_{\sss0})}+
  a_{\sss2})$$
and we are done.

\line{\hfil$\diamondsuit$}
\goodbreak
\bigskip

The approximate equality~(5.3)
extends to the derivatives of the functions
involved. The character $\Dif$ stands for the
Jacobian matrix.

\bigskip

{\bf Proposition 6.4 } \sl In the hypotheses of
Proposition~6.3, for each $\mu>0$ and for each
$\gamma >0$ there exists $d_{\sss0}\ge0$ such that
\rm
$$\left(\matrix{\bar p\in\D^\circ,\hfill&
  \hbox{dist}(\bar p,\partial\D)
  \ge\gamma \,,\hfill\cr
  \bar q\in q_{\sss1}+\D\,,\hfill&
  \hbox{dist}(\bar q,q_{\sss1}+\partial\D)
  \ge d_{\sss0}\hfill\cr}\right)\quad
  \Rightarrow\quad
  \Bigl\| \Dif p_{\sss\infty}(\bar p,\bar q)-
  \Dif\Pi(\bar p,\bar q)\Bigr\|\le\mu\,.$$

{\bf Proof. } Fix $\gamma >0$. For any
$d_{\sss0}\ge0$ such that
$$(A_{\sss1}e^{-\lambda_{\sss1} d_0})^{1/2}\le
  {\lambda_{\sss1}\gamma \over2}$$
we set $\varepsilon=(A_{\sss1}
e^{-\lambda_{\sss1} d_0})^{1/2}$. Applying
Lemma~6.2 we get that for all $\bar p \in\D$,
$\bar q\in q_{\sss1}+\D$ such that
$$\hbox{dist}\,(\bar q,q_{\sss1}+\partial\D)\ge
  -{1\over\lambda_{\sss1}}\ln{\varepsilon^2\over
   A_{\sss1}}=
  d_{\sss0}$$
and for all $t\ge0$, we have
$$\eqalign{
  \bigl|Dq(t,\bar p,\bar q)\bigr|&=
  \Bigl|\Pi R(t,0,\bar p,\bar q)
  {D\Pi^\prime\choose D\Pi}\Bigr|\le
  {1\over 2\varepsilon}
  (\varepsilon|D\Pi^\prime|+
  |D\Pi|)e^{\varepsilon t}+
  {1\over 2}|D\Pi^\prime|\le\cr
  &{}\le
  {1\over2\varepsilon}\Bigl(
  {\lambda_{\sss1}\gamma \over2}
  |D\Pi^\prime|+|D\Pi|\Bigr)
  e^{\lambda_{\sss1}\gamma 
  t/2}+{1\over2}|D\Pi^\prime|\,.\cr}$$
On the other hand, if moreover
$\hbox{dist}(\bar p,\partial\D)\ge\gamma$, we
have
$$\hbox{dist}\bigl(q(t,\bar p,\bar q),
  q_{\sss1}+\partial\D\bigr)\ge
  d_{\sss0}+\gamma  t\,,$$
so that
$\|\H\V(q(t,\bar p,\bar q))\|\le
A_{\sss1}e^{-\lambda_{\sss1}\gamma  
t-\lambda_{\sss1} d_0}$.
Putting the pieces together, and reminding that
$\varepsilon=(A_{\sss1}
e^{-\lambda_{\sss1} d_0})^{1/2}$:
$$\eqalign{
  \Bigl|Dp_{\sss\infty}(\bar p,\bar q)-
  D\Pi(\bar p,\bar q)\Bigr|&=
  \biggl|
  \int_0^{+\infty}\mskip-14mu
  -\H\V(q(t,\bar p,\bar q))\,Dq(t,\bar p,\bar q)
  \,dt\biggr|\le\cr
  &{}\le
  A_{\sss1}^{1/2}\Bigl({|D\Pi^\prime|\over2}+
  {|D\Pi|\over\lambda_{\sss1}\gamma }\Bigr)
  e^{-\lambda_{\sss1} d_0/2}+
  {A_{\sss1}|D\Pi^\prime|\over
  2\lambda_{\sss1}\gamma }
  e^{-\lambda_{\sss1} d_0}\,.\cr}$$
It is clear that we can choose $d_{\sss0}$ so
large that the last quantity is as small as we
wish.

\line{\hfil$\diamondsuit$}
\goodbreak

  \vfil\eject

%%%%%%%%%%%%%%%%%
\centerline{{\bfuno 7. First Order
             Differentiability}}
\smallskip
\centerline{{\bfuno with convexity on the
             potential}}
\bigskip

If we assume that the potential $\V$ is a {\it
convex} function, then the Hessian matrix $\H\V$
is nonnegative definite, and, from
equation~(6.3), 
$$\ddot z(t)=
  -\H\V(q(t,\bar p,\bar q))\,z(t)\,,
  \eqno(7.1)$$
it follows that $\ddot z\cdot z\le0$. This
lets us hope that we can derive a much sharper
estimate on $|Dq(t,\bar p,\bar q)|$ than the
mere less-than-exponential of Section~6.
We will also assume that the
quadratic form associated with $\H\V(q)$ behaves
monotonically with respect to $q$. Supposing $\V$
to be three times differentiable is not strictly
necessary, but will simplify the proofs.

\bigskip
\goodbreak

{\bf Hypothesis 7.1 } \sl $\V$ is a $C^3$
function and there exists $q_{\sss1}\in\Rfi^n$
such that
\item{ i)} $\V$ is convex on $q_{\sss1}+\D$;
\item{ii)} for all $q^\prime,\;q^{\prime
  \prime}\in q_{\sss1}+\D$ and all $z\in\Rfi^n$
  we have
  $$q^{\prime\prime}\in q^\prime+\D\quad
    \Rightarrow\quad
    \H\V(q^{\prime\prime})z\cdot z\le
    \H\V(q^\prime)
    z\cdot z\,\hbox{;}$$
\item{iii)} there exists a weakly decreasing
  function $h_{\sss1}\colon\Rfi_+\to\Rfi$ such
  that $\int_0^{+\infty}\mskip-7mu 
  xh_{\sss1}(x)dx<+\infty$ and\rm
  $$q\in q_{\sss1}+\D\quad\Rightarrow\quad
    \|\H\V(q)\|\le h_{\sss1}\Bigl(\,
    \hbox{dist}\bigl(q,q_{\sss1}+
    \partial\D\bigr)\,\Bigr)\,.$$

\bigskip
\goodbreak

{\bf Lemma 7.2 } \sl 
Suppose that Hypo\-theses~1.1, 4.1, 4.2 and~7.1
i), ii) hold. Then, for all $x,\,y\in\Rfi^n$ and
$t\ge0$ we have \rm
$$\Bigl(\bar p\in\D\,,\quad\bar q\in q_{\sss1}
  +\D\Bigr)\quad\Rightarrow\quad\left\{
  \matrix{|\Pi R(t,0,\bar p,\bar q)
  {x\choose y}|\le
  |x|+t\bigl(|y|+\|\H\V(q_{\sss1})\|^{1/2}\,
  |x|\bigr)\,,\hfill\cr
  |\Pi^\prime R(t,0,\bar p,\bar q)
  {x\choose y}|\le
  |y|+\|\H\V(q_{\sss1})\|^{1/2}\,|x|\,.
  \hfill\cr}\right.$$

\bigskip
\goodbreak

{\bf Proof. } Let $\bar p\in\D$, $\bar q\in
q_{\sss1}+\D$ be fixed. Let $z(t)=\Pi R(t,0,
\bar p,\bar q){x\choose y}$, and consider the
following Liapunov function:
$$L(t,\bar p,\bar q)=|\dot z(t)|^2+
  \H\V(q(t,\bar p,\bar q))z(t)\cdot z(t)\,.$$
$L$ is a nonnegative quantity because $\V$ is
convex in the points where the Hessian is
evaluated.  We are going to prove that $L$ is
decreasing in $t$ for $t\ge0$. Take the
derivative with respect to $t$, and remind
equation~(7.1): $$\eqalign{
  {d\over dt}L(t,\bar p,\bar q)&{}=
  2\dot z(t)\cdot 
  \ddot z(t)+ 2\H\V(q(t,\bar p,\bar q))
  z(t)\cdot\dot z(t)+
  \Bigl({d\over dt}\bigl(
  \H\V(q(t,\bar p,\bar q))\,\bigr)\Bigr)
  z(t)\cdot z(t)=\cr
  &{}=\lim_{s\searrow t}{1\over s-t}
  \bigl(\H\V(q(s,\bar p,\bar q))-
  \H\V(q(t,\bar p,\bar q))\bigr)
  z(t)\cdot z(t)
  \,.\cr}$$
This expression is nonpositive because of
Hypo\-thesis~7.1 ii) and because 
$q(s,\bar p,\bar q)\in q(t,\bar p,\bar q)+\D$ for
$s\ge t\ge0$. We thus have, for all $t\ge0$: 
$$|\dot z(t)|^2\le
  L(t,\bar p,\bar q)\le 
  L(0,\bar p,\bar q)\le
  |\dot z(0)|^2+\H\V(\bar q)z(0)\cdot z(0)\le
  |y|^2+\H\V(q_{\sss1})x\cdot x\,,$$
and so
$$\Bigl|\Pi^\prime R(t,0,\bar p,\bar q)
  {x\choose y}\Bigr|=|\dot z(t)|\le
  |y|+\|\H\V(q_{\sss1})\|^{1/2}\,|x|\,.$$
The other inequality comes from the last one and
from $|z(t)|\le|x|+\int_0^t|\dot z(s)|ds$.

\line{\hfil$\diamondsuit$}
\goodbreak
\bigskip

{\bf Proposition 7.3 } \sl Suppose that
Hypo\-theses~1.1, 4.1, 4.2 and~7.1 hold. Then
the asymptotic velocity
$p_{\sss\infty}$ is a $C^1$ function of the
initial data. \rm

\bigskip
\goodbreak

{\bf Proof. } Let $(\bar p_{\sss0},
\bar q_{\sss0})$, $\gamma >0$, $t_{\sss1}\in\Rfi$,
$U$ as in the proof of Proposition~6.3. Then,
for all $(\bar p,\bar q)\in U$ and $t\ge
t_{\sss1}$ we have
$$\|\H\V(q(t,\bar p,\bar q))\|\le
  h_{\sss1}(\gamma (t-t_{\sss1}))\,.$$
On the other hand, from Lemma~7.2 and
formula~(6.4), we get that, again for all  
$(\bar p,\bar q)\in U$, 
$t\ge t_{\sss1}$: 
$$|Dq(t,\bar p,\bar q)|\le a_{\sss1}+
  a_{\sss2}(t-t_{\sss1})\,,$$
where
$$a_{\sss1}\,:=\sup_{(\bar p,\bar q)\in U}
  |Dq(t_{\sss1},\bar p,\bar q)|\,,\quad
  a_{\sss2}\,:=\sup_{(\bar p,\bar q)\in U}
  \Bigl(
  |Dp(t_{\sss1},\bar p,\bar q)|
  +\|\H\V(q_{\sss1})\|^{1/2}\,
  |Dq(t_{\sss1},\bar p,\bar q)|
  \Bigr)\,.$$
We can finally write, for all $(\bar p,\bar
q)\in U$, $t\ge t_{\sss1}$:
$$\bigl|-\H\V(q(t,\bar p,\bar q))
  Dq(t,\bar p,\bar q)\bigr|\le
  \bigl(a_{\sss1}+a_{\sss2}(t-t_{\sss1})
  \bigr)\,
  h_{\sss1}(\gamma (t-t_{\sss1}))$$
and we are done.

\line{\hfil$\diamondsuit$}
\goodbreak
\bigskip

Also Proposition~6.4 remains true in the
modified hypotheses.

\nobreak
\bigskip

{\bf Proposition 7.4 } \sl In the hypotheses of
Proposition~7.3, for each $\mu>0$, and each
$\gamma >0$ there exists  $d_{\sss0}\ge0$ such 
that \rm
$$\left(\matrix{\bar p\in\D^\circ,\hfill&
  \hbox{dist}(\bar p,\partial\D)
  \ge\gamma \,,\hfill\cr
  \bar q\in q_{\sss1}+\D\,,\hfill&
  \hbox{dist}(\bar q,q_{\sss1}+\partial\D)
  \ge d_{\sss0}\hfill\cr}\right)\quad
  \Rightarrow\quad
  \Bigl\| \Dif p_{\sss\infty}(\bar p,\bar q)-
  \Dif\Pi(\bar p,\bar q)\Bigr\|\le\mu\,.$$

{\bf Proof. } Fix $\mu>0$ and $\gamma >0$. Let 
$\bar p\in\D^\circ$, dist$\,(\bar p,\partial\D)
\ge\gamma $, $\bar q\in q_{\sss1}+\D$ and
dist$\,(\bar q,q_{\sss1}+\partial\D)\ge
d_{\sss0}$. Then, for all $t\ge0$
$$\|\H\V(q(t,\bar p,\bar q))\|\le
  h_{\sss1}(\gamma  t+d_{\sss0})\,.$$
Moreover, from Lemma~7.2, we get
$$\eqalign{
  \bigl|Dq(t,\bar p,\bar q)\bigr|=
  \Bigl|\Pi R(t,0,\bar p,\bar q)
  {D\Pi^\prime\choose D\Pi}\Bigr|\;&{\le}\;
  |D\Pi^\prime|+t\Bigl(
  |D\Pi^\prime|+
  \|\H\V(q_{\sss1})\|^{1/2}\,|D\Pi|\Bigr):=\cr
  :&{=}\;b_{\sss1}+b_{\sss2}t\,.\cr}$$
In conclusion,
$$\eqalign{
  \bigl|Dp_{\sss\infty}(\bar p,\bar q)-
  D\Pi(\bar p,\bar q)\bigr|&=
  \Bigl|\int_0^{+\infty}\mskip-14mu
  -\H\V(q(t,\bar p,\bar q))
  Dq(t,\bar p,\bar q)\,dt\Bigr|\le\cr
  &{}\le\int_0^{+\infty}\mskip-14mu
  (b_{\sss1}+b_{\sss2}t)\,
  h_{\sss1}(\gamma  t+d_{\sss0})\,dt\,.\cr}$$
It is clear that we can choose $d_{\sss0}$ so
large that the last integral is as small as
needed.

\line{\hfil$\diamondsuit$}
\goodbreak
\bigskip

{\bf Proposition 7.5 } \sl In the hypotheses of
Proposition~5.3 and either~6.4 or~7.6, 
the mapping
$p_{\sss\infty}$ is surjective from
$\Rfi^n\times\Rfi^n$ onto $\D^\circ$.\rm

\bigskip

{\bf Proof. }
From Proposition~4.3, the image of
$p_{\sss\infty}$ is contained in $\D^\circ$.
To prove the reverse inclusion, let
$\bar p_{\sss0}\in\D^\circ$. We can solve the
equation 
$p_{\sss\infty}(\bar p,\bar q)=\bar p_{\sss0}$ 
via the contraction principle. Let
$\gamma :={}$dist$(\bar p_{\sss0},\partial\D)>0$.
From Proposition~5.3 and either~6.4 or~7.6, there
exists $\bar q\in\Rfi^n$ such that
$$\bigl|p_{\sss\infty}(\bar p,\bar q)-\bar p
  \bigr|\le{\gamma \over2}\,,\quad
  \bigl\|\Dif p_{\sss\infty}(\bar p,\bar q)-
  \Dif\Pi\bigr\|\le{1\over2}$$
for all $\bar p\in\D$ such that
dist$(\bar p,\partial\D)\ge\gamma /2$.
Now the mapping 
$\bar p\mapsto \bar p_{\sss0}+\bar p-
p_{\sss\infty}(\bar p,\bar q)$ 
is a contraction
of the closed ball $\{\bar p\in\Rfi^n\,:\,
|\bar p-\bar p_{\sss0}|\le\gamma /2\}$ 
into itself. The corresponding fixed point 
$\bar p$ solves 
$p_{\sss\infty}(\bar p,\bar q)=\bar p_{\sss0}$.
Actually, we could make it without
differentiability, in the mere hypotheses of
Propositions~5.2 and~5.3, if we were willing to
conjure up Brouwer's fixed point theorem.

\line{\hfil$\diamondsuit$}

   \vfil\eject

%%%%%%%%%%%%%%%%%
\centerline{{\bfuno 8. Higher Order
             Differentiability}}
\bigskip

Let us denote by $D_1$, $D_2$ the partial
derivative operators with respect to any two
components of $(\bar p,\bar q)$, by $D_{\sss1,2}$
the second order derivative $D_1D_2$, and by
$\Dif^3\V(q)$ the third differential of the
potential $\V$, regarded as a bilinear operator
from $\Rfi^n\times\Rfi^n$ into $\Rfi^n$, with
the norm
$$\|\Dif^3\V(q)\|:=\sup_{|x|\le1,\,|y|\le1}
  |\Dif^3\V(q)(x,y)|\,.$$

To get the second order differentiability of the
asymptotic velocity $p_{\sss\infty}$, what we
need is (local) uniform integrability of
$$\int_0^{+\infty}\biggl(-
  \underbrace{\Dif^3\V(q(t,\bar p,\bar q))
  \Bigl(D_1q(t,\bar p,\bar q),
  D_2q(t,\bar p,\bar q)\Bigr)}_{:=
  r(t,\bar p,\bar q)}-
  \H\V(q(t,\bar p,\bar q))
  D_{\sss1,2}q(t,\bar p,\bar q)\biggr)dt\,.
  \eqno(8.1)$$
To this purpose we must obtain an estimate on
the growth of $z(t)=D_{\sss1,2}q(t,\bar p,\bar q)$,
which is a solution of the non-homogeneous
linear differential equation:
$$\ddot z(t)=-\H\V(q(t,\bar p,\bar q))z(t)-
  r(t,\bar p,\bar q)\,,$$
or, in first order system form:
$${d\over dt}{z(t)\choose \dot z(t)}=
  \left(
  \matrix{0&I_n\cr
  -\H\V(q(t,\bar p,\bar q))&0\cr}
  \right)
  {z(t)\choose \dot z(t)}+
  {0\choose -r(t,\bar p,\bar q)}.$$
Remind the evolution operator $R$ introduced in
Section~6. The function $z(t)$ can be expressed
via $R$ with the classical method of variation
of the constants:
$${z(t)\choose\dot z(t)}=\int_0^t
  R(t,s,\bar p,\bar q)
  {0\choose-r(s,\bar p,\bar q)}ds
  \eqno(8.2)$$
(note that in our case $z(0)=\dot z(0)=0$).

We already have two sets of hypotheses that give
an estimate of the evolution operator. All we
are left to do is to give bounds on 
$r(t,\bar p,\bar q)$.

In the setting of Section~6, 
$D_1q(t,\bar p,\bar q)$ and 
$D_2q(t,\bar p,\bar q)$ grow less than
exponentially as $t\to+\infty$. If we assume
that $\|\Dif^3\V(q(t,\bar p,\bar q))\|$
decreases exponentially, then our scheme seems
to work out.

\bigskip
\goodbreak

{\bf Hypothesis 8.1 } \sl $\V$ is a $C^3$ 
function and there exist $q_{\sss2}\in\Rfi^n$,
$A_{\sss2}\ge0$, $\lambda_{\sss2}>0$ such that 
\rm 
$$q\in q_{\sss2}+\D\quad\Rightarrow\quad
  \|\Dif^3\V(q)\|\le A_{\sss2}\exp\Bigl(\,
  -\lambda_{\sss2}\,\hbox{dist}\bigl(
  q,q_{\sss2}+\partial\D\bigr)\,\Bigr).$$

\bigskip
\nobreak

We can safely assume that $q_{\sss2}$ coincides
with the $q_{\sss1}$ of Hypo\-thesis~6.1.

\goodbreak

In the frame of Section~7, 
$D_1q(t,\bar p,\bar q)$ and 
$D_2q(t,\bar p,\bar q)$ grow linearly as
$t\to+\infty$. Therefore the following
assumption seems appropriate.

\nobreak
\bigskip

{\bf Hypothesis 8.2 } \sl $\V$ is a $C^3$
function and there exist $q_{\sss2}\in\Rfi^n$
and a weakly decreasing function
$h_{\sss2}\colon\Rfi_+\to \Rfi$ such that \rm
$\int_0^{+\infty}x^2h_{\sss2}(x)dx<+\infty$ 
and\rm
$$q\in q_{\sss2}+\D\quad\Rightarrow\quad
  \|\Dif^3\V(q)\|\le h_{\sss2}\Bigl(\,
  \hbox{dist}\bigl(q,q_{\sss2}+\partial\D
  \bigr)\,
  \Bigr)\,.$$

\bigskip

Here, too, we can assume $q_{\sss2}$ to coincide
with the $q_{\sss1}$ of Hypo\-thesis~7.1.

\bigskip
\goodbreak

{\bf Proposition 8.3 } \sl Suppose that
Hypo\-theses~1.1, 4.1, 4.2, 6.1 and~8.1 hold. Then
the asymptotic velocity $p_{\sss\infty}$ is a
$C^2$ function of the initial data.\rm

\bigskip
\goodbreak

{\bf Proof. } Let $(\bar p_{\sss0},
\bar q_{\sss0})$, $\gamma >0$, $U$,
$\varepsilon>0$, $t_{\sss1}\le t_{\sss0}$
be as in the proof of Proposi\-tion~6.3. 
We showed there that
$$|D_iq(t,\bar p,\bar q)|\le a_{\sss1}e^{
  \varepsilon(t-t_0)}+a_{\sss2}\,,\qquad
  i=1,2$$
for all $t\ge t_{\sss0}$ and all $(\bar p,\bar q)
\in U$. This, together  with Hypo\-thesis~8.1
shows the uniform integrability of the first
half of the integral~(8.1):
$$|r(t,\bar p,\bar q)|\le 
  A_{\sss2}e^{-\lambda_{\sss2}\gamma 
  (t-t_1)}\bigl(a_{\sss1}e^{\varepsilon(t-t_0)}
  +a_{\sss2})\le a_{\sss3}e^{(\varepsilon-
  \lambda_{\sss2}\gamma )t}\,.$$
We must now estimate $z(t)$:
$${z(t)\choose\dot z(t)}=
  {D_{\sss1,2}q(t_{\sss0},\bar p,\bar q)\choose
  D_{\sss1,2}p(t_{\sss0},\bar p,\bar q)}+
  \int_{t_0}^tR(t,s,\bar p,\bar q)
  {0\choose-r(s,\bar p,\bar q)}ds\,.
  \eqno(8.3)$$
We have 
$$R(t,s,\bar p,\bar q)=R\bigl(
  t-s,0,p(s,\bar p,\bar q),q(s,\bar p,\bar q)
  \bigr)\,.
  \eqno(8.4)$$
From Lemma~6.2 we get a constant $a_{\sss4}$
such that
$$\Bigl|\Pi R(t,s,\bar p,\bar q)
  {0\choose y}\Bigr|\le a_{\sss4}|y|
  e^{\varepsilon(t-s)}$$
for all $(\bar p,\bar q)\in U$, $t_{\sss0}\le
s\le t$. Hence
$$\biggl|\int_{t_0}^t\Pi R(t,s,\bar p,\bar q)
  {0\choose-r(s,\bar p,\bar q)}ds\biggr|\le
  a_{\sss3}a_{\sss4}\int_{t_0}^t
  e^{\varepsilon(t-s)}e^{
  (\varepsilon-\lambda_{\sss2}
  \gamma )s}ds\le a_{\sss5}e^{\varepsilon t}\,.$$
Estimating $|D_{\sss1,2}q(t_{\sss0},\bar p,\bar q)|$
on $U$ by a constant, we can write $|z(t)|\le
a_{\sss6}e^{\varepsilon t}$. The last step is:
$$\bigl|-\H\V(q(t,\bar p,\bar q))D_{\sss1,2}
  q(t,\bar p,\bar q)\bigr|\le
  A_{\sss2}e^{-\lambda_{\sss2}\gamma 
  (t-t_1)}a_{\sss6}e^{
  \varepsilon t},$$
for all $(\bar p,\bar q)\in U$, $t\ge t_{\sss0}$,
and the proof is complete.

\line{\hfil$\diamondsuit$}
\goodbreak
\bigskip

{\bf Proposition 8.4 } \sl Suppose that
Hypo\-theses~1.1, 4.1, 4.2, 7.1 and~8.2 hold. 
Then the asymptotic velocity $p_{\sss\infty}$ is 
a $C^2$ function of the initial data. \rm

\bigskip

{\bf Proof. } Let $(\bar p_{\sss0},
\bar q_{\sss0})$, $U$, $\gamma $, $t_{\sss1}$ as
in the proof of Proposition~6.3. In the proof of
Proposition~7.3 we saw that 
$$\bigl|D_iq(t,\bar p,\bar q)\bigr|\le
  a_{\sss1}+a_{\sss2}(t-t_{\sss1})\,,
  \qquad i=1,2$$
for all $(\bar p,\bar q)\in U$, $t\ge t_{\sss1}$.
This, together with Hypo\-thesis~8.2, gives 
$$|r(t,\bar p,\bar q)|\le\bigl(
  a_{\sss1}+a_{\sss2}(t-t_{\sss1})\bigr)\,
  h_{\sss2}\bigl(\gamma (t-t_{\sss1})\bigr)\,,$$
and half of the job is done. Using again
formulas~(8.3) and (8.4) (with $t_{\sss1}$
instead of $t_{\sss0}$), together with
Lemma~7.2, we get
$$\eqalign{
  \biggl|\int_{t_1}^t\Pi^\prime R(t,s,
  \bar p,\bar q){0\choose-r(s,\bar p,\bar q)}ds
  \biggr|&\le\int_{t_1}^t|r(s,\bar p,\bar q)|\,ds
  \le\cr
  &{}\le\int_{t_1}^{+\infty}\mskip-14mu
  \bigl(a_{\sss1}+a_{\sss2}
  (s-t_{\sss1})\bigr)\,
  h_{\sss2}\bigl(\gamma (s-t_{\sss1})
  \bigr)\,ds:= a_{\sss3}\,,\cr}$$
$$|\dot z(t)|\le a_{\sss4}+a_{\sss3}:=
  a_{\sss5}\,,\qquad
  |z(t)|\le a_{\sss6}+a_{\sss5}(t-t_{\sss1})\,.$$
Finally:
$$\bigl|-\H\V(q(t,\bar p,\bar q))D_{\sss1,2}
  q(t,\bar p,\bar q)\bigr|\le
  \bigl(a_{\sss6}+a_{\sss5}(t-t_{\sss1})\bigr)\,
  h_{\sss2}\bigl(\gamma (t-t_{\sss1})\bigr)$$
for all $(\bar p,\bar q)\in U$, $t\ge t_{\sss1}$.
The proof is complete.

\line{\hfil$\diamondsuit$}
\goodbreak
\bigskip

The hypotheses that guarantee higher order
derivatives of $p_{\sss\infty}$  are now easy to
guess. Denote by $\Dif^m\V$  the $m$-th
differential of $\V$, viewed as a multilinear
operator form  $(\Rfi^n)^{m-1}$ into $\Rfi^n$,
endowed with the norm
$$\|\Dif^m\V(q)\|:=\sup\Bigl\{
  \bigl|\Dif^m\V(q)(x^{(1)},\ldots,x^{(m-1)})
  \bigr|\;\colon\;
  x^{(i)}\in\Rfi^n,\; |x^{(i)}|\le1\;\Bigr\}\,.$$

\bigskip

{\bf Hypothesis 8.5 } \sl

\nobreak

\item{$ H_m$)} $\V$ is a $C^{m+1}$ 
     function and
     there exist $\lambda_m>0$, $A_m>0$ and
     $q_m\in\Rfi^n$ such that
     $$q\in q_m+\D\quad\Rightarrow\quad
       \|\Dif^{m+1}\V(q)\|\le A_m\exp\Bigl(\,
       -\lambda_m\,\hbox{\rm dist}
       \bigl(q,q_m+\partial\D\bigr)\,
       \Bigr)\,\hbox{;}$$
\medskip
\item{$ H_m^\prime$)} 
  $\V$ is a $C^{m+1}$ function and
  there exist $q_m\in\Rfi^n$ and a weakly
  decreasing function 
  $h_m\colon\Rfi_+\to\Rfi$ such that
  $\int_0^{+\infty}\mskip-4mu
  x^mh_m(x)dx<+\infty$
  and
  $$q\in q_m+\D\quad\Rightarrow\quad
    \|\Dif^{m+1}\V(q)\|\le h_m\Bigl(\,
    \hbox{\rm dist}\bigl(q,q_m+\partial\D\bigr)\,
    \Bigr)\,.$$
\rm

\bigskip

The following proposition can be proved by
induction on $m$, with essentially the same
reasoning used in Propositions~8.3 and~8.4.

\nobreak
\bigskip

{\bf Proposition 8.6 } \sl The asymptotic
velocity $p_{\sss\infty}$ is a $C^m$ function of
the initial data, $m\ge2$,
if we assume Hypo\-theses~1.1,
4.1, 4.2, and either~i) or~ii) of the following:
\medskip

\item{i)} Hypothesis~6.1 plus $ H_2$, 
  $H_3$,
  $\ldots$ , $H_m$ 
  of Hypo\-thesis~8.5;

\item{ii)} Hypo\-thesis~7.1 plus 
  $H_2^\prime$,
  $H_3^\prime$, $\ldots$ , 
  $H_m^\prime$ of
  Hypo\-thesis~8.5.
\rm
   \vfil\eject

%%%%%%%%%%%%%%%%%
\centerline{{\bfuno 9. Complete
                       Integrability
                       and Persistence}}
\bigskip

It is time to exploit the regularity theory
developed so far to achieve our main goal:
the integrability of the system
$$\dot q={\partial\Ham\over\partial p}\,,\quad
  \dot p=-{\partial\Ham\over\partial q}\,,\qquad
  \Ham(p,q):={1\over2}|p|^2+\V(q)\,.
  \eqno(9.1)$$

We introduce the notation $X_f$ to mean the
Hamiltonian vector field defined by some smooth
$f\colon\Rfi^{2n}\to\Rfi$. In particular
$X_{\Ham}$ is the vector field in~(9.1). We will
say that $X_f$ is {\it complete} if all the
solutions of the Hamilton equations 
$\dot q=\partial f/\partial p$,
$\dot p=-\partial f/\partial q$
are global, that is, defined on the whole 
of $\Rfi$.

We are going to prove that the components
of the asymptotic velocity are $n$ first 
integrals independent and in involution. Moreover
we can include the Hamiltonian $\Ham$ into a set
of $n$ globally independent first integrals in
involution $\F_1,\ldots,\F_n$, whose associated
vector fields  $X_{\F_1},\ldots,X_{\F_n}$ are
complete. The $\F_i$, $2\le i\le n$, will be
obtained from $p_{\sss\infty}$ through a linear
transformation. We will use the fact that, in our
hypotheses, ${1\over2}|p_{\sss\infty}|^2$ is
just $\Ham$, up to an immaterial additive 
constant (Corollary~4.6).

\bigskip
\goodbreak

{\bf Theorem 9.1 (Complete Integrability)} \sl 
Assume the hypotheses of Proposition either~8.3
or~8.4. Then the $n$ components
$$p_{\sss\infty,1}\,,\ldots,p_{\sss\infty,n}$$
of the asymptotic velocity are independent 
$C^2$ integrals of motion and they are (pairwise)
in involution. This means that, for all 
$(\bar p,\bar q)\in\Rfi^n\times\Rfi^n$, 
the gradients  
$$\nabla p_{\sss\infty,1}(\bar p,\bar q)
  \,,\;\nabla p_{\sss\infty,2}(\bar p,\bar q)\,,
  \ldots,\; \nabla p_{\sss\infty,n}(\bar p,\bar q)
  \eqno(9.2)$$
are linearly independent 
and the Poisson brackets vanish identically:
$$\{p_{\sss\infty,i}\,,\,p_{\sss\infty,j}\}
  (\bar p,\bar q)=0\,.$$
Hence the system~(9.1) is integrable by 
quadratures.

Furthermore, there exists an orthogonal
transformation  $A\colon\Rfi^n\to\Rfi^n$ such 
that the functions $\F_1,\ldots,\F_n $ defined as
$$\eqalign{&\F_1:=\Ham\,\cr
  &\F_i:=(Ap_{\sss\infty})_i \quad\hbox{for }
  i=2,3,\ldots,n\,,\cr}
  \eqno(9.3)$$ 
are independent $C^2$ integrals of
motion, (pairwise) in involution and the
Hamiltonian vector fields $X_{\F_i}$ ($1\le i\le
n$) are complete.

\noindent
In particular, the system~(9.1) is completely
integrable.\rm

\bigskip
\goodbreak

{\bf Proof. }
Claiming that the gradients in~(9.2) are
independent is the same as saying that the
Jacobian matrix $\Dif p_{\sss\infty}$ 
($p_{\sss\infty}$ thought of as column vector) 
has maximum rank. Denote by 
$\{\Phi^t\}_{t\in\Rfi}$ the flow of the 
system~(1.1), i.e., $\Phi^t(\bar p,\bar q)=
\bigl(p(t,\bar p,\bar q),
q(t,\bar p,\bar q)\bigr)$. 
Since $p_{\sss\infty}$ is an
integral of motion, 
$$p_{\sss\infty}\circ\Phi^t=p_{\sss\infty}
  \qquad\forall t\in\Rfi\,.$$
By differentiating at $(\bar p,\bar q)$ we have
$$\Dif p_{\sss\infty}
  \bigl(\Phi^t(\bar p,\bar q)\bigr)
   \Dif\Phi^t(\bar p,\bar q)=
  \Dif p_{\sss\infty}(\bar p, \bar q)\,.$$
From this we have that 
$$\hbox{rank }\Dif p_{\sss\infty}$$ 
is a constant of motion, because 
$\Dif\Phi^t(\bar p,\bar q)$ is invertible. 

\noindent From Poisson's Theorem (see e.g.~[1],
Section~40) we know that the Poisson brackets
$$\{p_{\sss\infty,i}\,,\,p_{\sss\infty,j}\}$$
are also constants of motion.  

\noindent We now use Proposition either 6.5
or~7.6. Along  any trajectory, the velocity
$p(t,\bar p,\bar q)$ eventually enters $\D^\circ$
and keeps from its boundary a distance larger 
than $\gamma >0$. Moreover,  $q(t,\bar p,\bar q)$
enters all the sets of the form $q_{\sss0}+\D$
(Proposition~4.4).
Hence, along any trajectory the derivatives of
$p_{\sss\infty}$ tend to the derivatives of the
projection $\Pi$ and we can compute: 
$$\eqalignno{
  \hbox{rank }\Dif p_{\sss\infty}(\bar p,\bar q)
  &{}=\hbox{rank }
  \Dif p_{\sss\infty}\bigl(\Phi^t(\bar p, \bar q)
  \bigr)= 
  \lim_{t\to+\infty}
  \hbox{rank }
  \Dif p_{\sss\infty}\bigl(\Phi^t(\bar p, \bar
  q)\bigr)=\cr
  &{}= \hbox{rank }\Dif \Pi=n\cr
\noalign{\hbox{(the set of the $n\times2n$
matrices with maximum rank is open in
$\Rfi^{2n^2}$) and}}
  \{p_{\sss\infty,i}\,,\,,p_{\sss\infty,j}\}
  (\bar p,\bar q)& = 
  \{p_{\sss\infty,i}\,,\,p_{\sss\infty,j}\}
  (\Phi^t(\bar p, \bar q))=
  \lim_{t\to+\infty}
  \{p_{\sss\infty,i}\,,\,p_{\sss\infty,j}\}
  (\Phi^t(\bar p, \bar q))=\cr
  & = \{\Pi_{\sss i}\,,\,\Pi_{\sss j}\}=0\cr}$$

\goodbreak
\medskip

Now, let us prove the second part of the theorem.
We can use the Remark~2.6 to construct an 
orthogonal transformation 
$A:\Rfi^n\to\Rfi^n$ (we will write $A$ also for
the associated matrix) such that the first
component $(Aw)_1$ is strictly positive for all
$w\in\D^\circ$. Hence, from Proposition~4.3 
we have that 
$$\bigl(A
  p_{\sss\infty}(\bar p,\bar q)\bigr)_1>0
  \qquad \forall (\bar p, \bar q)\,. 
  \eqno(9.4)$$
Define $\F_1,\F_2,\ldots,\F_n$ as in~(9.3).
Proposition~4.5 implies that
$\F_1=\Ham={1\over2}|p_{\sss\infty}|^2+\inf\V$. 
Let $\Dif\F$ be the Jacobian matrix of
$\F:=(\F_1,\ldots,\F_n)^T$ (the $^T$ means
transposition of matrices). Proving that
$\F_1,\ldots,\F_n$ are independent is equivalent
to proving that the rank of $\Dif\F$ is $n$. 
Since $\F$ is a first integral of motion, the
rank of $\Dif\F$ does not change along each
trajectory.%
         {\parfillskip=0pt\par\goodbreak
          \parskip=0pt\noindent}%
From Proposition either~6.5 or~7.6 we know that
along all trajectories the matrix  $\Dif
p_{\sss\infty}$ tends to the matrix
$\Dif\Pi=(I_n,0)$, so that
$\Dif(Ap_{\sss\infty})\to\Dif(A\Pi)=(A,0)=
(I_n,0){A\choose0}$, and   $\Dif\F_1$ tends to
$(p_{\sss\infty}^T,0)=((A^TAp_{\sss\infty})^T,0)=
((Ap_{\sss\infty})^T,0){A\choose0}$. 
Thus we can write 
$$\eqalign{
  \hbox{rank }&\Dif \F(\bar p,\bar q)=
  \lim_{t\to+\infty}\hbox{rank }
    \Dif \F(\Phi^t(\bar p,\bar q))=\cr
  \noalign{\medskip}
  &{}=
  \hbox{rank}
  \left(
   \matrix{
    \left.
     \matrix{
     (Ap_{\sss\infty})_1 &&\ldots&&
        (Ap_{\sss\infty})_n\cr
     \hfil\hfil\hfil0\hfil&1&0&\ldots&
            \hfil0\hfil\hfil\hfil\hfil\cr
     \hfil\hfil\hfil0\hfil&0&1&\ldots&
            \hfil0\hfil\hfil\hfil\hfil\cr
     \hfil\hfil\hfil\vdots\hfil&\vdots&&\ddots&
            \hfil\vdots\hfil\hfil\hfil\hfil\cr
     \hfil\hfil\hfil0\hfil&0&0&\ldots&
            \hfil1\hfil\hfil\hfil\hfil\cr
}
    \right|
  &
   \matrix{
       &\vdots&\cr
       \cdots&0&\cdots\cr
       &\vdots&\cr                        }
  \cr}
  \right)
  {A\choose0}=n\cr}
 $$
because of~(9.4). So $\F_1,\ldots,\F_n $ are
independent.

\noindent
We have
$$\{\Ham\,,\,\F_i\}=\{\Ham\,,\,
  (Ap_{\sss\infty})_i\}=
 \sum_{j=1}^n A_{ij}\{\Ham\,
 ,\,p_{\sss\infty,j}\}=0\,$$ 
because the Poisson brackets are bilinear.
Therefore, $\F_1,\ldots,\F_n $ are pairwise in
involution since 
$p_{\sss\infty,1}\,,\ldots,\allowbreak
p_{\sss\infty,n}$ 
are.

\noindent
The last result in particular says that $\Ham$ 
is a  first integral of each 
$X_{\F_i}\,$. 
This implies the completeness
of each vector field $X_{\F_i}\,$ 
as we can see by the
same argument which gave the completeness of 
$X_\Ham$ in Section~1.

\line{\hfill$\diamondsuit$}
\goodbreak
\bigskip

We remark that we could have included the
Hamiltonian $\Ham$ into a set of $n$ {\it
locally} independent integrals of motion,
extracted from $p_{\sss\infty}$, in the following
way. We could have simply chosen a nonvanishing
component of $p_{\sss\infty}$ (there is always
one, locally, because $p_{\sss\infty}\in\D^\circ$)
and replaced it with $\Ham$. This  would give
independence generally only in a neighbourhood of
each level set of $p_{\sss\infty}$, unless we
have been lucky or deft from the very beginning
in the choice of the orthonormal reference basis
of $\Rfi^n$.

\goodbreak
\medskip

It is legitimate to ask what happens to the
integrability of our system if we perturb the
potential $\V$ in a compact set $K$:
$$H(p,q):={1\over2}|p|^2+\V(q)+f(q)\,,
  \eqno(9.5)$$
with $f$ a smooth function vanishing outside $K$.
The new potential $\V+f$ does not need to be a 
cone potential, or, even if it is, it may not 
have the same cone. Think for example to the case
when the cone $\C$ of $\V$ has empty interior. 
The global hypo\-theses~4.1 and~4.2 are very
sensitive to the cones $\C$ and $\D$, and there
is no hope to verify them for the new cone.
Nevertheless, we will prove that if $|\nabla f|$
is sufficiently small, then all the trajectories
of the new system eventually quit $K$ for good,
the potential $\V$ leads henceforth undisturbed,
and all our conclusions about regularity and
integrability remain true.

\bigskip
\goodbreak

{\bf Theorem 9.2 (Persistence) } \sl
Suppose that $\V$ verifies the hypotheses of
Theorem~9.1. Let $K\subset\Rfi^n$ be compact. 
Then there exists an $\varepsilon>0$ with the
following property. If $f\colon\Rfi^n\to\Rfi$ is
a $C^3$ function with support in $K$ and 
$$\sup|\nabla f|<\varepsilon\,,
  \eqno(9.6)$$
then the Hamiltonian system whose Hamiltonian 
$H$ is given by Equation~(9.5) is completely
integrable. Namely, all trajectories have 
asymptotic velocities for which all the claims 
in Theorem~9.1 apply.\rm

\bigskip
\goodbreak

{\bf Proof. } We still denote by $\C$ and $\D$ 
the cones associated with $\V$. Let
$v\in\C\backslash\{0\}$. Since $K$ is compact and
$\D$ has nonempty interior, there exist 
$q^\prime$, $q^{\prime\prime}\in\Rfi^n$,
$q^{\prime\prime}\in q^\prime+\D$, such that
$$K\subset q^\prime+\D\,,\qquad
  q\cdot v\le q^{\prime\prime}\cdot v\quad
  \forall q\in K\,.$$
Define $\varepsilon$ by
$$\varepsilon:=\inf\bigl\{
  -\nabla\V(q)\cdot v\;:\;
  q\in q^\prime+\D\,,\;
  q\cdot v\le q^{\prime\prime}\cdot v\bigr\}\,.$$
From Hypothesis~4.2, $\varepsilon$ is positive.
Suppose that $f$ is a $C^3$ function
with support in $K$ and verifies~(9.6). Denote by
$(p(t,\bar p,\bar q),q(t,\bar p,\bar q))$ the
trajectories corresponding to the new
Hamiltonian~(9.5). We claim that
$$\forall(\bar p,\bar q)\quad\exists
  t_{\sss0}\in\Rfi\quad\hbox{such that}\quad
  \forall t\ge t_{\sss0}\quad
  q(t,\bar p,\bar q)\cdot v>q^{\prime\prime}
  \cdot v\,.
  \eqno(9.7)$$
In fact, there certainly exists 
$\tilde q\in\Rfi^n$ such that 
$q(t,\bar p,\bar q)\in\tilde q+\D$ for all
$t\in\Rfi$ because Hypo\-thesis~4.1 still holds
for $\V+f$ and~$\D$. We can safely assume
$q^\prime\in\tilde q+\D$. Let
$$\varepsilon^\prime:=\inf
  \bigl\{-\nabla\V(q)\cdot v\;:\;
  q\in\tilde q+\D\,,\;
  q\cdot v\le q^{\prime\prime}\cdot v\bigr\}\,.$$
We can write
$$\Bigl(\; q\in\tilde q+\D\quad\hbox{and}\quad
  q\cdot v\le q^{\prime\prime}\cdot v\;\Bigr)
  \quad\Rightarrow\quad
  -\nabla(\V+f)(q)\cdot v\ge
  \min\{\varepsilon-\inf|\nabla f|\,,\,
  \varepsilon^\prime\}>0\,.$$
Now the same reasoning as in the proof of
Proposition~4.3 leads to~(9.7). 

\noindent To conclude, we only need to note that
the $t_{\sss0}$ in~(9.7) can be chosen locally
independent of $(\bar p,\bar q)$, and that in all
the hypotheses on $\V$ after Section~4 we can
always suppose $q_{\sss0}\,$, $q_{\sss1}\,\ldots$
to belong to $q^{\prime\prime}+\D$.

\line{\hfil$\diamondsuit$}
\goodbreak

  \vfil\eject

%%%%%%%%%%%%%%%%%
\centerline{{\bfuno 10. Examples}}

\bigskip

We are going to present a class of
examples for which our assumptions for complete
$C^2$ integrability hold, in the convex case. 
Conditions for $C^k$ integrability, $2\le
k\le+\infty$, are easily added.

\bigskip

{\bf Hypotheses 10.1 } \sl
For $\alpha=1,\ldots ,N$, let $f_\alpha$ be a
$C^3$ real function, defined either on all of
$\Rfi$ or on the interval $]0,+\infty[$, such
that, for all $\alpha$,
$$\hbox{{\rm inf}}\, f_\alpha=0\,,  \quad 
  \hbox{{\rm sup}}\,
  f_\alpha=+\infty\,.\eqno(10.1)$$
$$f^\prime_\alpha(x)<0\quad\forall x\,,
  \eqno(10.2)$$
$$f_\alpha^{\prime\prime}(x)>0\quad
  \forall x\ge1\,,
  \eqno(10.3)$$
$$f_\alpha^{\prime\prime\prime}
  \hbox{ is weakly decreasing on }
  [1,+\infty[\,,\quad
  f_\alpha^{\prime\prime\prime}(x)<0\quad
  \forall x\ge1\,.
  \eqno(10.4)$$
Let $v_{\sss1},\ldots, v_{\sss N}\in\Rfi^n
\backslash\{0\}$ be such  that 
$$v_\alpha\cdot v_\beta\ge0 \quad\forall
   \alpha\,, \beta\,.\eqno(10.5)$$
\rm

\goodbreak
\bigskip

Define the following potential
$\V$:
$$\V(q):=\sum_{\alpha=1}^N f_\alpha(q\cdot
  v_\alpha)\,.
  \eqno(10.6)$$
either on $\Rfi^n$ or in the set
$\{ w\in\Rfi^n\;:\;
w\cdot v_\alpha>0\quad
\forall\alpha=1,\ldots, N\}\,.$
The gradient, Hessian and third differential of
$\V$ are given by 
$$\eqalignno{
  \nabla\V(q)&=\sum_{\alpha=1}^N
  f_\alpha^\prime(q\cdot v_\alpha)\;v_\alpha\,,
  &(10.7)\cr
  \H\V(q)&=\sum_{\alpha=1}^N
  f_\alpha^{\prime\prime}
  (q\cdot v_\alpha)\;v_\alpha\otimes
  v_\alpha&(10.8)\cr
  \Dif^3\V(q)&=\sum_{\alpha=1}^N
  f_\alpha^{\prime\prime\prime}
  (q\cdot v_\alpha)\;
  v_\alpha\otimes v_\alpha\otimes v_\alpha\,,
  &(10.9)\cr}$$
where $\otimes$ indicates the tensor product:
$(v_\alpha\otimes v_\alpha)_{i,j}:=
v_{\alpha,i}v_{\alpha,j}$, etc.

\goodbreak

From~(10.1), (10.2), (10.5) and~(10.7) it is clear
that Hypothesis~1.1 is satisfied for~$\V$.
Also the other requirements for integrability
are met, and the proof will be made in several
steps, culminating in the statement of
Proposition~10.3 and its corollaries.

\goodbreak

It will be convenient to have from the start a
$q_{\sss0}\in\Rfi^n$ such that 
$$q_{\sss0}\cdot v_\alpha\ge1\quad\forall\alpha
  \eqno(10.10)$$
(for example,  $q_{\sss0}:=\rho\sum_{\beta=1}^N
v_\beta$, for $\rho$ large).
Recall Proposition~2.5 and note that, for all
$q\in q_{\sss0}+\D$, $\alpha_{\sss0}=1,\ldots,N$
$$q\cdot v_{\alpha_0}=(q-q_{\sss0})\cdot 
  v_{\alpha_0}
  +q_{\sss0}\cdot v_{\alpha_0}\ge
  \min_\beta|v_\beta|\,\hbox{dist}(q,q_{\sss0}
  +\partial\D)+1\,.
  \eqno(10.11)$$

\bigskip
\goodbreak

{\bf Lemma 10.2 } \sl The closure of the
convex cone $\C$ generated by the forces
$-\nabla\V$ coincides with the convex cone
generated by the $v_\alpha\,$: 
$$\bar\C=\biggl\{\sum_{\alpha=1}^N c_\alpha
  v_\alpha\;\colon\;
  c_\alpha\ge0\biggr\}\,,$$
and the dual $\D$ of $\C$ is given by 
$\{w\,:\,w\cdot v_\alpha\ge0\;\forall\alpha\}$.
\rm

\bigskip
\goodbreak

{\bf Proof. } Denote by $\bar C$ the cone
generated by $v_1,\ldots,v_N$ and by $D$ its
dual. Of course $\bar\C\subset\bar C$, so that
$D\subset\D$. Take $w\in\partial D$. We are done
if we show that $w\in\partial\D$. Consider the
following two sets of indices: $I_1:=\{
\alpha\,:\, w\cdot v_\alpha>0\}$,
$I_2:=\{\beta\,:\,w\cdot v_\beta=0\}$. From
Proposition~2.5 and the fact that
$w\in\partial D$ we deduce that
$I_2\ne\emptyset$.
Evaluate now $-\nabla\V$ along the line
$q_{\sss0}+\tau w$ ($q_{\sss0}$ given by~(10.10)),
for $\tau\to+\infty$; the terms
$f_\alpha^\prime((q_{\sss0}+\tau w)\cdot
v_\alpha)$ vanish for $\alpha\in I_1$, so that
$$\C\ni-\nabla\V(q_{\sss0}+\tau w)
  \;\to\;
  -\sum_{\beta\in I_2} f_\beta^\prime(q_{\sss0}
  \cdot v_\beta)\,v_\beta:=\tilde v\in\bar\C\,.$$
We have $\tilde v\ne0$ because
$-f_\alpha^\prime>0$ and the $v_\beta$ lie in the
interior of a half-space (the cone $\bar C$ is
proper). From the definition of $I_2$ we see that
$w\cdot\tilde v=0$ and finally $w\in\partial\D$
because of Proposition~2.4. The formula for $\D$
is an easy consequence.

\line{\hfil$\diamondsuit$}
\bigskip
\goodbreak

{\bf Lemma 10.3 } \sl Let $g\colon[0,+\infty[\to
\Rfi$ be a nonnegative, weakly decreasing
function such that
$$\int_0^{+\infty}\mskip-14mu
  x^mg(x)\,dx<+\infty$$
for some integer $m\ge0$. Then
$\lim_{x\to+\infty}x^{m+1}g(x)=0$. \rm

\bigskip
\goodbreak

{\bf Proof. } Suppose there exists
$x_i\nearrow+\infty$ such that
$x_i^{m+1}g(x_i)\ge\varepsilon>0$ for
all~$i\ge1$. Then
$$x_{i-1}<x\le x_i\quad\Rightarrow\quad
  x^mg(x)\ge x^mg(x_i)\ge\varepsilon{x^m\over
  x_i^{m+1}}\,.$$
Define $h\colon[0,+\infty[\to\Rfi$ as
$$h(x):=\cases{0&if $0\le x\le x_1$,\cr
  \varepsilon x^m/x_i^{m+1}&if $x_{i-1}<x\le
  x_i$, $i\ge2$.\cr}$$
Then $h(x)\le x^mg(x)$ and
$$\int_0^{+\infty}\mskip-14mu h(x)\,dx=
  \sum_{i=2}^{+\infty}{\varepsilon\over
  x_i^{m+1}}\cdot{x_i^{m+1}-x_{i-1}^{m+1}\over
  m+1}=
  {\varepsilon\over m+1}\sum_{i=2}^{+\infty}
  \biggl(1-{x_{i-1}^{m+1}\over x_i^{m+1}}
  \biggr).$$
We can safely assume, for instance, that
$x_i\ge2x_{i-1}$ for all $i\ge2$, and this
yields~$h$, and hence $x^mg(x)$ too, to have
infinite integral, against the hypothesis.

\line{\hfil$\diamondsuit$}
\bigskip
\goodbreak

{\bf Lemma 10.4 } \sl Let $f\in
C^M([0,+\infty[)$, $M\ge1$, and suppose that,
for all $x\ge0$, $m=0,\ldots,M$,
$$f^{(m)}(x)\cases{>0&if $m$ is even,\cr
  <0&if $m$ is odd.\cr}$$
Then $\int_0^{+\infty}x^{m-1}|f^{(m)}(x)|\,dx
<+\infty$ for all $m=1,\ldots,M$. \rm

\goodbreak
\bigskip

{\bf Proof. } It is certainly true if $M=1$.
Suppose it is true for~$M-1$ and write
$$\int_0^{+\infty}\mskip-14mu
  x^{M-1}f^{(M)}(x)\,dx=
  x^{M-1}f^{(M-1)}(x)|_{x=0}^{+\infty}-
  (M-1)\int_0^{+\infty}\mskip-14mu
  x^{M-2}f^{(M-1)}(x)\,dx\,.$$
The last integral converges for the induction
hypothesis. The term $x^{M-1}f^{(M-1)}(x)$ is
infinitesimal as $x\to+\infty$ again because
$x^{M-2}f^{(M-1)}(x)$ is integrable, with the
help of Lemma~10.3.

\line{\hfil$\diamondsuit$}
\bigskip
\goodbreak

{\bf Verification of 4.1 } Let $M\ge0$. From 10.1
and 10.2 we get $x_{\sss M}\in\Rfi$ such that
$$f_\alpha(x)\ge M\qquad \forall x\le x_{\sss M}
  \quad
  \forall \alpha=1,\ldots ,N\,.$$
Let $q_{\sss M}\in\Rfi^n$ such that
$$q_{\sss M}\cdot v_\alpha\le x_{\sss M}\quad
  \forall\alpha$$
(for example, $q_{\sss M}=\theta\sum_\alpha
v_\alpha$ for $\theta$ negatively large). Then,
for all  $q\in\Rfi^n\backslash(q_{\sss M}+\D)$
there exists $\alpha_{\sss M}$ such that
$q_{\sss M}\cdot  v_{\alpha_M}\le x_{\sss M}$ and
hence 
$$\V(q)\ge f_{\alpha_M}(q\cdot v_{\alpha_M})
  \ge f_{\alpha_M}(x_{\sss M})\ge M\,.$$

\line{\hfil$\diamondsuit$}
\bigskip
\goodbreak

{\bf Verification of 4.2 } Let
$q^\prime$ belong to the domain of $\V$ and
$q^{\prime\prime} \in q^\prime +\D$. Let
$v\in\bar\C\backslash\{0\}$.
Up to a reordering of indices, we can write 
$$v=\sum_{\beta=1}^{N^\prime} c_\beta v_\beta
  \in\bar\C \qquad \hbox{with }1\le N^\prime
  \le N \hbox{ and }
  c_\beta>0\,.\eqno(10.12)$$
By~(10.7), (10.12), (10.2) and~(10.5)
we have
$$\eqalign{
  -\nabla\V(q)\cdot v &=
  -\sum_{\alpha=1}^N
     f_\alpha^\prime (q\cdot v_\alpha)\,
  v_\alpha\cdot v =
  \sum_{\alpha=1}^N
  \sum_{\beta=1}^{N^\prime}
    c_\beta \bigl(-f_\alpha^\prime
  (q\cdot v_\alpha)\bigr)\, 
     v_\alpha\cdot v_\beta \ge \cr
  &\ge -\sum_{\beta=1}^{N^\prime} 
  c_\beta 
  f_\beta^\prime
  (q\cdot v_\beta)|v_\beta|^2\,.\cr}$$
Moreover
$$q\in q^\prime +\D\quad\Rightarrow\quad
  \forall\beta\quad
  q^\prime \cdot v_\beta\le q\cdot v_\beta\,.$$
On the other hand
$$q\cdot v\le q^{\prime\prime} \cdot v
  \quad\Rightarrow\quad
  \exists\beta_{\sss0} \hbox{ such that }
  q\cdot v_{\beta_0}\le 
  q^{\prime\prime} \cdot v_{\beta_0}\,.$$
Hence
$$\Bigl(\; q\in q^\prime +\D\hbox{ and }
  q\cdot v\le q^{\prime\prime} \cdot v\;
  \Bigr)\quad\Rightarrow\quad
  \exists\beta_{\sss0}\hbox{ such that }
  q^\prime \cdot v_{\beta_0}\le
  q\cdot v_{\beta_0}\le
  q^{\prime\prime} \cdot v_{\beta_0}\,.$$
For each $\beta=1,\ldots,N^\prime$, define
$$\varepsilon_{_{\scriptstyle\beta}}:=
  \min\bigl\{ -c_\beta f_\beta^\prime(x)\;\colon
  \; q^\prime \cdot v_\beta\le x\le
  q^{\prime\prime} \cdot v_\beta\bigr\}>0\,.$$
We can conclude that
$$\Bigl(\; q\in q^\prime +\D\quad\hbox{and}\quad
  q\cdot v\le q^{\prime\prime} \cdot v\;\Bigr)
  \quad\Rightarrow\quad
  -\nabla\V(q)\cdot v\ge
  \min_\beta \varepsilon_{_{\scriptstyle\beta}}
  >0\,.$$

\line{\hfil$\diamondsuit$}
\bigskip
\goodbreak

{\bf Verification of 5.1 }  
Recall $q_{\sss0}$ from formula~(10.9).
Since $x\mapsto|f_\alpha^\prime(x)|$ is
weakly decreasing on $[1,+\infty[$
and from~(10.11), we can
compute, for all $q\in q_{\sss0}+\D$,
$$|\nabla\V(q)| \le
  \sum_{\alpha=1}^N |f_\alpha^\prime
  (q\cdot v_\alpha)|\;|v_\alpha|\le
  h_{\sss0}\Bigl(\,\hbox{dist}
  \bigl(q,q_{\sss0}+\partial\D\bigr)\,
  \Bigr)\,,$$
where
$$h_{\sss0}(x):=\sum_{\alpha=1}^N|v_\alpha|\;
  \bigl| f_\alpha^\prime(\min_\beta|v_\beta|
  x+1)\bigr|\,,$$
and $h_{\sss0}$ is weakly decreasing and
integrable on  $[0,+\infty[$ because each 
$|f_\alpha^\prime|$ is weakly decreasing and
integrable on $[1,+\infty[$.

\line{\hfil$\diamondsuit$}
\bigskip
\goodbreak

{\bf Verification of 7.1 }i). From Formula~(10.3)
we see that each $f_\alpha$ is convex on
$[1,+\infty[$. So the potential $\V$ is convex
on $q_{\sss0}+\D$, $q_{\sss0}$ given by~(10.10),
because it is sum of convex functions.

\line{\hfil$\diamondsuit$}
\bigskip
\goodbreak

{\bf Verification of 7.1 }ii). From Formula
(10.4) we see that $f_\alpha^{\prime\prime}$ is a
weakly decreasing function on $[1,+\infty[$. If
$q^\prime, q^{\prime\prime}\in q_{\sss0}+\D$, 
$q^{\prime\prime}\in q^\prime+\D$, then we have
$q^{\prime\prime}\cdot v_\alpha\ge
q^\prime\cdot v_\alpha\ge q_{\sss0}\cdot v_\alpha
\ge1$ and
$$f_\alpha^{\prime\prime}(q^{\prime\prime}\cdot
  v_\alpha)\le f_\alpha^{\prime\prime}
  (q^\prime\cdot v_\alpha)\,.$$
Hence, from~(10.8),
$$\eqalign{
  \H\V(q^{\prime\prime})z\cdot z&=
  \sum_{\alpha=1}^Nf_\alpha^{\prime\prime}
  (q^{\prime\prime}\cdot v_\alpha)(v_\alpha\cdot
  z)^2\le\cr
  &{}\le\sum_{\alpha=1}^Nf_\alpha^{\prime\prime}
  (q^\prime\cdot v_\alpha)(v_\alpha\cdot z)^2=
  \H\V(q^\prime)z\cdot z\,.\cr}$$

\line{\hfil$\diamondsuit$}
\bigskip
\goodbreak

{\bf Verification of 7.1 }iii). We have, from
Formula~(10.8) and~(10.11), for all $q\in
q_{\sss0}+\D$, 
$$\|\H\V(q)\|\le
  \sum_{\alpha=1}^N\bigl|f_\alpha^{\prime\prime}
  (q\cdot v_\alpha)\bigr|\;|v_\alpha|^2\le
  h_{\sss1}\Bigl(\,\hbox{dist}
  \bigl(q,q_{\sss0}+\partial\D\bigr)\,
  \Bigr)\,,$$
where
$$h_{\sss1}(x):=\sum_{\alpha=1}^N
  |v_\alpha|^2\bigl|f_\alpha^{\prime\prime}
  (\min_\beta|v_\beta|x+1)\bigr|\,,$$
and $h_{\sss1}$ is weakly decreasing on
$[0,+\infty[$ and $\int_0^{+\infty}\mskip-4mu
x|h_{\sss1}(x)|\,dx<+\infty$ because of
Lemma~10.4.

\line{\hfil$\diamondsuit$}
\bigskip
\goodbreak

{\bf Verification of 8.2 } From~(10.9) and~(10.4)
we have, for all $q\in q_{\sss0}+\D$,
$$\|\Dif^3\V(q)\|\le h_{\sss2}\Bigl(
  \hbox{dist}\bigl(q,q_{\sss0}+\partial\D\bigr)
  \Bigr)\,,$$
where
$$h_{\sss2}(x):=\sum_{\alpha=1}^N
  |v_\alpha|^3\bigl|f_\alpha^{\prime\prime\prime}
  (\min_\beta|v_\alpha|x+1)\bigr|$$
and, as usual, $h_{\sss2}$ is weakly decreasing on
$[0,+\infty[$ and $\int_0^{+\infty}\mskip-4mu
x^2h_{\sss2}(x)\,dx<+\infty$.

\line{\hfil$\diamondsuit$}
\bigskip
\goodbreak

{\bf Proposition 10.5 } \sl Suppose that the
functions $f_1,\ldots,f_N$ and the vectors
$v_1,\ldots, v_N$ satisfy the Hypotheses~10.1.
Let the potential $\V$ be defined by
Formula~(10.6). Then Theorem~9.2 applies, so that
the Hamiltonian system
$$\dot q=p\,,\quad\dot p=-\nabla\V(q)$$
is $C^2$-completely integrable.

\noindent If we assume, moreover, that the
$f_\alpha$ are $C^{k+1}$, $2<k\le+\infty$, and
that for all $4\le m\le k+1$, $\alpha=1,\ldots,N$,
$x\ge1$ we have 
$$f_\alpha^{(m)}(x)\;\cases{>0&
  if $m$ is even\cr
  <0& if $m$ is odd,\cr}$$
(and that $|f^{(k+1)}|$ be weakly decreasing if
$k<+\infty$),
then $p_{\sss\infty}$ is $C^k$.
\rm

\bigskip
\goodbreak

{\bf Corollary 10.6 } \sl Let $v_1,\ldots,v_N\in
\Rfi^n\backslash\{0\}$ be such that $v_\alpha
\cdot v_\beta\ge0$ for all $\alpha,\beta$.
Let $r>0$ and define the potential
$$\V(q):=\sum_{\alpha=1}^N
  {1\over(q\cdot v_\alpha)^r}$$
on the set $\D^\circ=\{q\in\Rfi^n\;\colon\;
q\cdot v_\alpha>0\;\forall\alpha\}$.
Then the associated Hamiltonian system is
$C^\infty$-completely integrable.
\rm

\goodbreak
\bigskip

{\bf Corollary 10.7 } \sl Let $v_1,\ldots,v_N\in
\Rfi^n\backslash\{0\}$ be such that $v_\alpha
\cdot v_\beta\ge0$ for all $\alpha,\beta$, and
let $c_\alpha>0$. Define the potential
$$\V(q):=\sum_{\alpha=1}^N
  c_\alpha\, e^{-q\cdot v_\alpha}$$
on $\Rfi^n$.
Then the associated Hamiltonian system is
$C^\infty$-completely integrable.\rm

 \vfil\eject

%%%%%%%%%%%%%%%%%
\centerline{{\bfuno 11. References}}
\bigskip \frenchspacing

\item{[1]} Arnold, V. I. (1978). 
   {\bf Mathematical methods in classical
   mechanics.} 
   Springer Verlag, Berlin.

\medskip

\item{[2]} Arnold, V. I. (ed.) (1988). 
   {\bf Encyclopaedia of mathematical sciences 3,
     Dynamical Systems III.} 
   Springer Verlag, Berlin.

\medskip

\item{[3]} Calogero, F. (1971). 
   {\it Solutions of the  one
   dimensional  $n$-body problems
   with quadratic and/or
   inversely quadratic pair potentials}.
   {\bf J. Math. Phys. 12}, pp.~419--436.

\medskip

\item{[4]}
   Calogero, F., \& Marchioro, C. (1974) 
   {\it Exact solution of a one-dimensional 
   three-body scattering problem with  
   two-body and/or three-body 
   inverse-square potentials}. 
   {\bf J. Math. Phys. 15}, 
   pp.~1425--1430.

\medskip

\item{[5]} Gutkin, E. (1985). 
   {\it Integrable  Hamiltonians with
   exponential potentials}. 
   {\bf Physica~D~16},
   pp.~398--404,
   North Holland, Amsterdam.

\medskip

\item{[6]}  Gutkin, E. (1985). 
   {\it Asymptotics of  trajectories 
   for cone potential}. 
   {\bf Physica~D~17},
   pp.~235--242.

\medskip 

\item{[7]}  Gutkin, E. (1987). 
   {\it  Continuity of scattering data 
   for particles on the line 
   with directed repulsive interactions}.
   {\bf J.~Math. Phys. 28}, 
   pp.~351--359.

\medskip

\item{[8]} Marchioro, C. (1970). 
   {\it Solution of a  three-body 
   scattering problem in one dimension}. 
   {\bf J.~Math. Phys.~11}, 
   pp.~2193--2196.

\medskip

\item{[9]} Moauro, V., \& Negrini, P. (1989).
  {\it On the inversion of Lagrange-Dirichlet
  theorem}. {\bf Differ. Integ. Equat.} to appear.

  \medskip

\item{[10]} Moser, J. (1975). 
   {\it Three integrable Hamiltonian 
   systems connected with isospectral
   deformations}. 
   {\bf Advances in Math.~16}, 
   pp.~197--220.

\medskip 

\item{[11]} Moser, J. (1983). 
   {\it Various aspects  of integrable
   Hamiltonian systems}. 
   In {\bf Dynamical Systems}
   (C.I.M.E. Lectures, Bressanone 1978),
   pp.~233--290, sec. print.,
   Birkh\"auser, Boston.

\medskip

\item{[12]} Oliva, W.M., 
    \& Castilla M.S.A.C. (1988).
    {\it On a class of $C^k$-integrable 
    Hamiltonian systems}. 
    Preprint \bf RT--MAP,
    \rm S\~ao Paulo.

\medskip 

\item{[13]} Yoshida, H., Ramani, A., 
    Grammaticos, B. \& Hietarinta, J. (1987). 
    {\it On the non-integrability of some 
    generalized Toda lattices}.
    {\bf Physica~144 A}, 
    pp.~310--321.

\medskip

\item{[14]} Ziglin, S.L. (1982, 1983). 
     {\it Branching  of solutions
     and non existence of first integrals 
     in Hamiltonian mechanics~I, II}. 
     {\bf Functional Anal. Appl.~16, 17}, 
     pp.~181--189, pp.~6--16.

\bigskip 
\centerline{\hbox to3cm{\hrulefill}}

%%%%%%%%%%%%%%%%%
               \vfil\eject\end

\end